\documentclass[%
 reprint,
 superscriptaddress,
 amsmath,amssymb,
 aps,
]{revtex4-2}
\usepackage{graphicx}
\usepackage{dcolumn}
\usepackage{bm}
\usepackage{wrapfig}
\usepackage{mhchem}
\usepackage{amssymb}
\usepackage{amsmath}
\usepackage{supertabular}
\usepackage[table,xcdraw]{xcolor}
\begin{document}

\title{How Can We Engineer Electronic Transitions Through Twisting and Stacking in TMDC Bilayers and Heterostructures? A First-Principles Approach}

\author{Yu-Hsiu Lin}

 \affiliation{Department of Chemical Engineering \& Materials Science, Michigan State University, East Lansing, Michigan 48824, United States.}
\author{William P. Comaskey}
 \affiliation{Department of Health Physics, University of Nevada, Las Vegas, Las Vegas, NV 89154, USA.}
 \affiliation{National High Magnetic Field Laboratory, Tallahassee, FL 32310, USA.}

\author{Jose L. Mendoza-Cortes}
 \email{jmendoza@msu.edu}
\affiliation{Department of Chemical Engineering \& Materials Science, Michigan State University, East Lansing, Michigan 48824, United States.}
\affiliation{Department of Physics, Michigan State University, East Lansing, Michigan 48824, United States.}

\date{\today}

\begin{abstract}
Layered two-dimensional (2D) materials exhibit unique properties not found in their individual forms, opening new avenues for material exploration. This study examines MX$_2$ transition metal dichalcogenides (TMDCs), where M is Mo or W, and X is S, Se or Te. These materials are foundational for the creation of hetero- and homo-bilayers with various stacking configurations. Recent interest has focused on twisted homogeneous bilayers, as critical twist angles can significantly alter material properties. This work highlights MX$_2$ TMDC bilayers with twisted angles that form Moiré patterns, essential to understanding the behaviors of these materials. We performed first-principles calculations using Density Functional Theory (DFT) with range-separated hybrid functionals on 30 combinations of six MX$_2$ materials with two stacking configurations, revealing that the building blocks and stacking arrangements influence the stability of the heterostructure and the band gap energy (E$_g$). In particular, the MoTe$_2$/WSe$_2$ heterostructure, shifted by 60\textdegree, exhibits a direct band gap, indicating potential for novel applications. Our investigation of homobilayers included fully relaxed and low-strain scenarios, examining various stacking styles and twisting angles. Under low-strain conditions, MoS$_2$, WS$_2$, and WSe$_2$ can exhibit direct or indirect band gaps at specific twist angles. Additionally, MoS$_2$ can transition between semiconductor and conductor states, showcasing diverse electronic properties. Critical twist angles, specifically 17.9\textdegree~and its corresponding angles (42.1\textdegree~, 77.9\textdegree~and 102.1\textdegree~), in twisted WS$_2$ and WSe$_2$ bilayers create symmetric Moiré patterns, leading to direct band gaps. The magnitude of the band gap energy can be tuned by varying the twist angles, which also affect the flatness of the electronic band. Like conventional stacking, most twisted TMDC bilayers exhibit favorable interlayer interactions but with more tailorable characteristics. Using heterostructures and controlled twist angles is a powerful approach in material engineering, enabling the manipulation of various electronic behaviors in advanced materials.
\end{abstract}

\maketitle
\begin{figure*}
 \centering
 \includegraphics[width=0.95\linewidth]{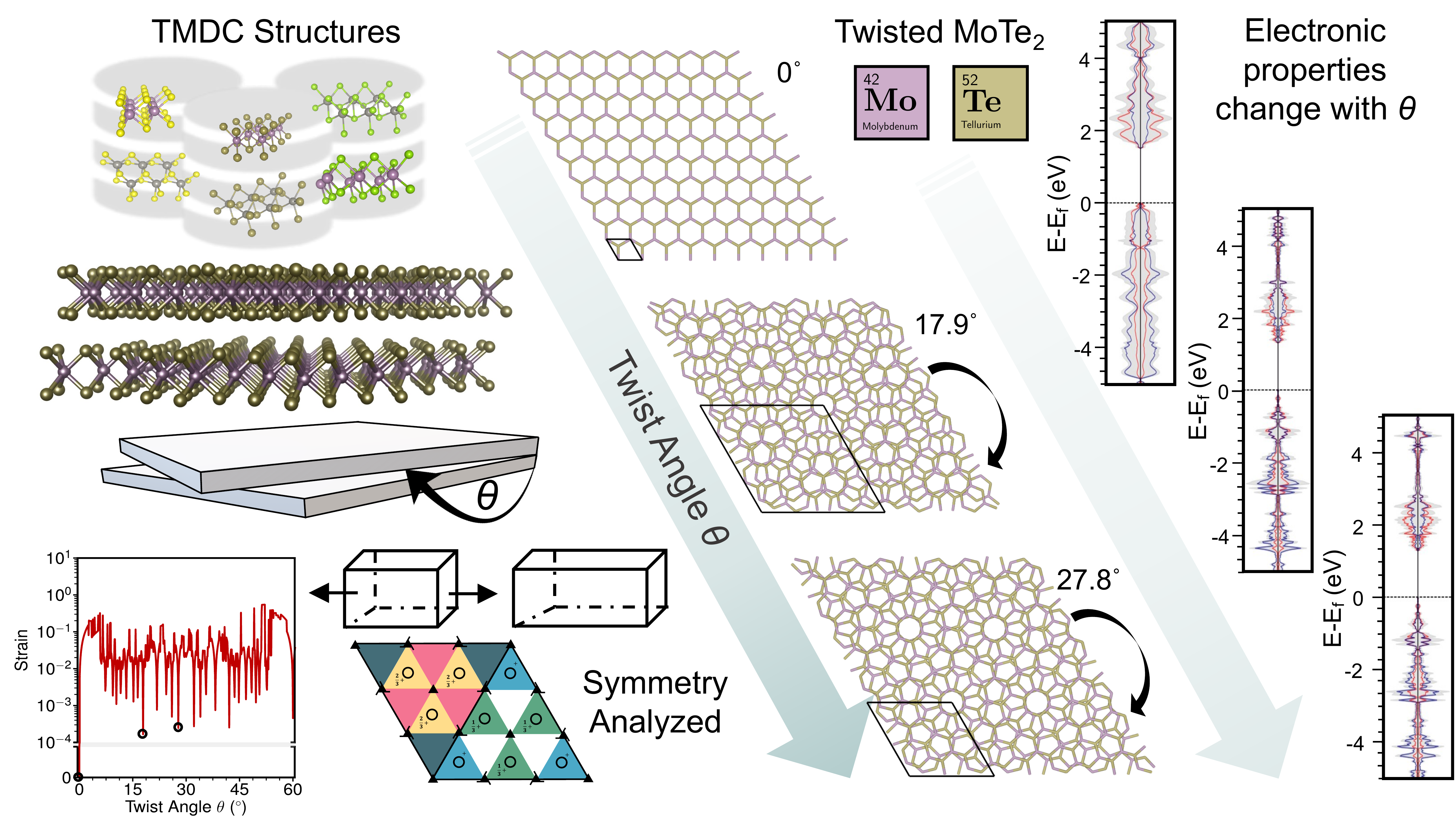}
 \caption{Workflow diagram illustrating the computational study of transition metal dichalcogenides. Starting from the initial structural information (in this case \ce{MoTe2}), the workflow progresses by inputting primitive cells into supercell-forge (a modified version of the supercell-core program). Selected moderate-sized structures undergo symmetry analysis to ensure computational tractability, followed by submission to an ab initio electronic structure workflow utilizing \textsc{Crystal23} for detailed analysis.}
 \label{fig:All}
\end{figure*}

\begin{figure*}
\centering
  \includegraphics[width=\textwidth]{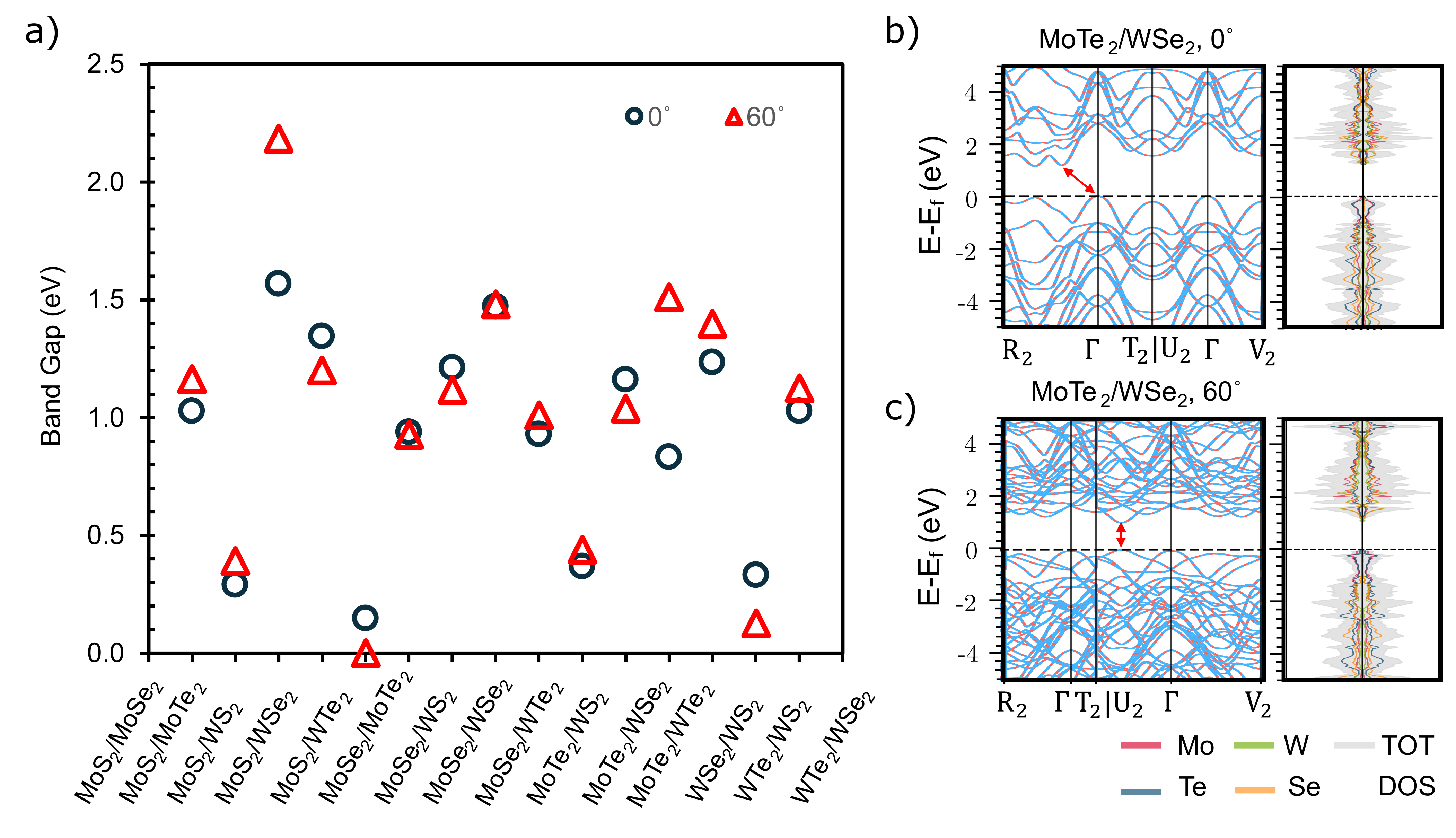}
  \caption{Comparative analysis of band gap shown for 15 bilayer TMDC heterostructures (a), each evaluated at either a 0 or 60-degree shift. In this plot, the 1H phase of MoS$_2$ is utilized, whereas Figure S2 incorporates the 1T phase of MoS$_2$, highlighting the variations in band gap due to different structural phases of the same material. Band structures and the density of states of MoTe$_2$/WSe$_2$ bilayers (b) without rotation, and (c) with rotation. The inserted arrows pinpoint the top of the valence band and the bottom of the conduction band.}
  \label{fig:1}
\end{figure*}

\section{Introduction}
Layered two-dimensional (2D) materials exhibit unique characteristics distinct from their bulk counterparts\cite{voiry2015phase}. The presence of weak van der Waals (vdW) interactions between these layers allows stacking in non-equilibrium configurations, unveiling a plethora of fascinating phenomena\cite{long2019robust, shi2018interlayer}. This versatility is not limited to heterostructures alone; homogeneous 2D materials are also at the forefront of scientific inquiry because of their ability to manipulate a wide range of properties. The field, abundant with a diverse array of building blocks and customization strategies, still holds many secrets yet to be uncovered. In navigating this vast landscape, first-principles calculations emerge as a crucial tool, providing efficient and insightful guidance in ongoing research within this domain.

Within the vast landscape of van der Waals (vdW) heterostructures, a variety of layered materials such as graphene, boron nitride (BN)\cite{dean2010boron}, transition metal dichalcogenides (TMDCs)\cite{sahoo2018one,jin2019observation}, and metal carbides and nitrides (MXenes) have been extensively utilized as fundamental building blocks. In this context, even simple bilayer configurations of TMDCs emerge as a robust platform for engineering applications given their shared properties and symmetries. In particular, in certain bilayer TMDC heterostructures, the generation of direct E$_g$ has been observed, paving the way for groundbreaking advancements in optoelectronics\cite{terrones2014bilayers, wang2016interlayer}. 
The role of symmetry is pivotal in determining the electronic characteristics of these materials. Transitioning TMDCs from their bulk form to 2D structures often involves breaking the inversion symmetry, which can lead to fascinating changes in the structure of the band, including the appearance of spin degeneracy\cite{jones2014spin, sung2020broken, xiao2012coupled}. Moreover, the hybridization of the density of states from different layers fosters interlayer charge transfer. This charge transfer plays a crucial role in the creation of an electric dipole moment, significantly contributing to the unique electronic properties of these heterostructures\cite{li2017binary}.

\begin{figure*}
 \centering
 \includegraphics[width=\textwidth]{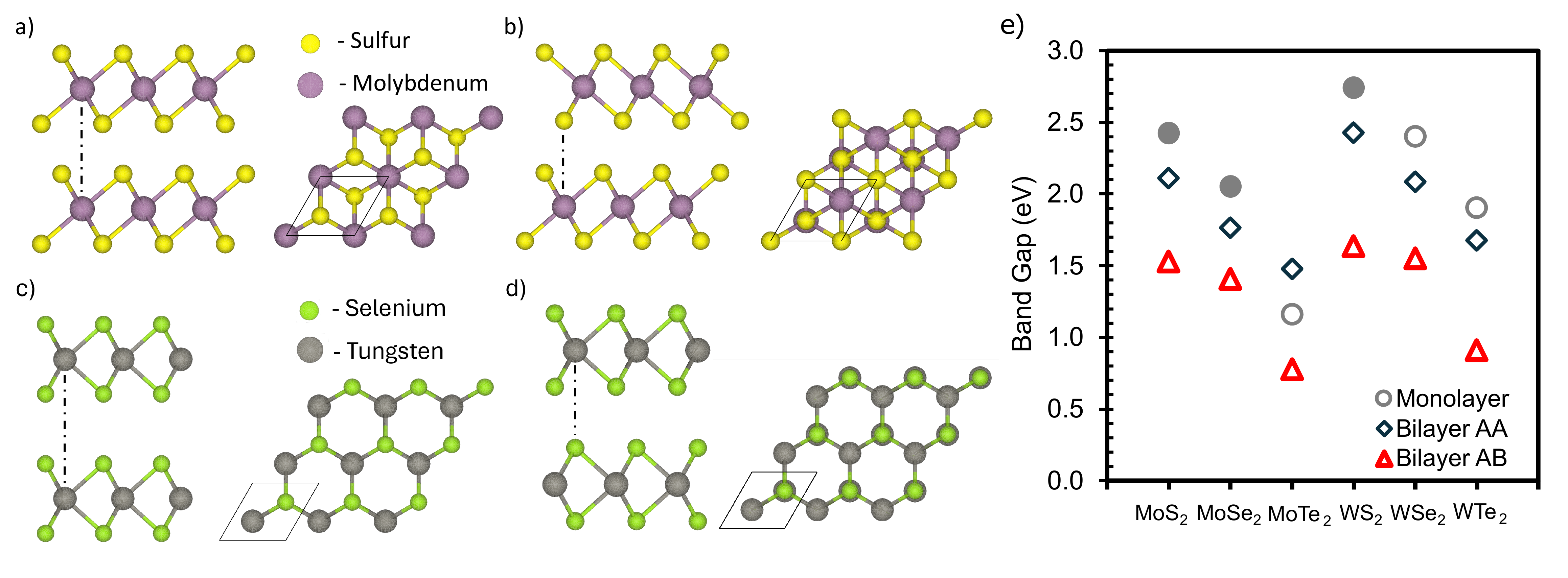}
 \caption{Top and side views of transition metal dichalcogenides (TMDCs) bilayers are shown. Panels (a-b) display the A-A and A-B stacking arrangements in the 1T phase of TMDCs, while panels (c-d) depict the A-A and A-B stacking in the 1H phase of TMDCs. Panel (e) shows the comparison of E$_g$ for six fully optimized TMDCs, showcasing their properties in monolayer form as well as in A-A and A-B stacked bilayer configurations. Hollow symbols are used to indicate structures with an indirect E$_g$, and filled symbols represent those with a direct E$_g$. It's noteworthy that this plot utilizes 1H-MoS$_2$, in contrast to Figure S3, which presents the corresponding data for the 1T phase of MoS$_2$, thus providing a comparative understanding of the influence of structural phases on E$_g$.}
 \label{fig:3}
\end{figure*}

Beyond heterostructures, homogeneous layered transition metal dichalcogenides (TMDCs) have been shown to display a wide range of electronic behaviors, heavily influenced by their geometric arrangements\cite{zhang2019transition, kim2012high}. For example, bilayer WS$_2$ has been observed to alternate between direct and indirect E$_g$ depending on its stacking configuration. However, comprehensive studies exploring the full gamut of TMDC material families across various stacking geometries remain limited. Additionally, interlayer coupling in 2D layered materials is sensitive not only to stacking types but also to twisted angles\cite{tang2021recent,lin2023recent}. 

The discovery of exceptional superconductivity in twisted bilayer graphene (TBG) at a specific ``magic angle'' of 1.1\textdegree~ has sparked significant interest in the study of twisted bilayer structures\cite{cao2018unconventional,cao2018correlated}. Since then, this research has expanded, exploring the topological and electronic properties of TBG and similar systems\cite{wu2019topological,yankowitz2019tuning,kim2017tunable}. At critical twist angles, symmetric superlattices can form, often characterized by Moir\'e patterns which create spatial confinement and enhance interlayer coupling. This leads to a flattened band structure and modulates the electronic properties, a phenomenon also observed in double-bilayer graphene\cite{bistritzer2011moire,cao2020tunable}. 

Given their hexagonal symmetry and van der Waals (vdW) interlayer interactions, TMDCs are postulated to offer similar opportunities for property manipulation through twisting of homobilayers\cite{liu2014evolution,puretzky2016twisted}. The optoelectronic behavior of MoS$_2$, for example, has been closely linked to its twist angles\cite{huang2014probing}. Unlike TBG, where flat bands occur within a narrow range of angles, TMDCs are expected to exhibit flat bands across a wider spectrum of twisted angles\cite{wang2020correlated,morell2010flat,zhang2020flat}. This makes TMDCs particularly versatile for engineering band structures. Therefore, twisting the layered configurations of homogeneous TMDC bilayers presents a promising approach to effectively modify their intrinsic properties.

Twisted TMDC bilayers demonstrate distinct electronic behaviors based on moiré angle. Early experiments \cite{Regan2020-po,Tang2020-pm} established that smaller moiré patterns exhibit band structure modifications without topological properties. Understanding these simpler systems is crucial as they provide a pathway to engineering topological states through twist angle manipulation. At larger periodicities ($\theta \approx 1-4^\circ$), the systems develop rich topological phenomena, including fractional Chern insulators and quantum spin Hall states \cite{Park2023-an,Kang2024-ig,Ji2024-uy,Kang2024-wj,Park2023-xg}. This angle-dependent transition from trivial to topological states makes TMDCs a versatile platform for studying and controlling electronic topology. We constrain the system size to ensure accurate treatment of long-range exchange interactions while maintaining computational feasibility to limit the size of the moiré lattices and avoid the accumulation of electron correlation errors inherent in range-separated hybrid functionals like HSE06, particularly due to self-interaction errors and the inability to fully satisfy the generalized Koopmans condition in certain systems \cite{Holmes2015-jb,ivady2013role}.

In this study, MX$_2$ transition metal dichalcogenides (TMDCs), where M is Mo or W and X is S, Se, or Te, were selected as the foundational elements for the construction of hetero-bilayers and homo-bilayers, with the aim of uncovering novel phenomena. A significant observation is that most bilayers demonstrate thermodynamically favorable interlayer interaction. This work delves into how the choice of building blocks influences such van der Waals interaction. Initially, we analyzed the combinations of these TMDCs in different stacking configurations, focusing on their E$_g$ and band structures. This was followed by a comparison between two types of stacking in homobilayers and their corresponding monolayers, setting the stage for a discussion on twisted homobilayers. In the context of heterobilayers, it was found that the selection of building blocks and stacking types critically influences the electronic properties. A notable discovery was the direct E$_g$ in the MoTe$_2$/WSe$_2$ heterobilayer. For homobilayers, while stacking types present a method to adjust properties, twisting emerges as a more dynamic variable for engineering interlayer couplings. Within specific ranges of twisted angles, homobilayers exhibiting symmetric Moir\'e patterns can induce regular shifts in E$_g$, toggle between types of E$_g$, and affect the flatness of electronic bands. This research paves the way for further exploration of innovative materials, with potential applications in energy conversion, electronics, and optoelectronics, marking a significant advancement in the field of material science.

\section{Results and discussion}

\subsection{Outline}

The foundational elements of this research are succinctly summarized in Figure~\ref{fig:All}, which sketches how the qualified materials are built, screened, and simulated. In detail, six TMDCs were first selected as the building blocks of the bilayers. A high-throughput coding was implemented to construct numerous preliminary bilayers by continuously rotating angles and applying strain to fit minimum unit cells. Due to structural stability, the need to maintain accurate electron correlation, and the limitation of computational power, we collected the preliminary bilayers with low intrinsic strain, a computable number of atoms, or high symmetry. In particular, employing range-separated hybrid functionals places increased demands on computational resources; thus, selecting moderately sized moiré structures helps ensure both computational tractability and sufficient accuracy of electron correlation. The geometries and electronic properties of the final candidates were computed using a hybrid DFT formulation utilizing range-separated hybrid functionals. The ensuing discussion is divided into different categories of layered materials. These include conventionally stacked hetero- and homo-bilayers, as well as twisted homo-bilayers.
\subsection{Hetero-bilayers}

In this work, we systematically explored TMDC heterostructures as a result of their remarkable material properties and structural versatility. Using six MX$_2$ TMDCs (M = Mo, W; X = S, Se, Te) as building blocks, we formed 15 bilayer heterostructures. These were further categorized into two stacking types: 0\textdegree~ and 60\textdegree~ shifts, resulting in 30 unique structures. Each structure was subjected to comprehensive geometric optimization and property analysis.
Stability, a key factor, was assessed using equation~\eqref{eqn:cohesive}, which calculates the binding energy by comparing the energy of the bilayers with those of their constituent monolayers. According to Figure S1, most heterobilayers exhibit a negative binding energy, indicating a greater energetic favorability than separate monolayers. In particular, WTe$_2$ bilayers tend to have a positive binding energy, suggesting an unfavorable interlayer interaction. In contrast, bilayers with MoTe$_2$ show stronger binding, which highlights how the choice of building blocks significantly influences stability.
The stacking method slightly alters the energy, as seen in the band gap analysis of these heterostructures in Figure~\ref{fig:1}(a). Although the band gaps vary with stacking in all heterostructures, those with 1T-MoS$_2$ consistently exhibit conductive properties, as Figure S2 illustrates, underscoring the impact of bonding types on electronic properties. Remarkably, MoTe$_2$/WSe$_2$ can toggle between indirect and direct band gaps (~1.04 eV). The band gap of 2D materials can transition between indirect and direct as a result of variations in their stacking configurations. This phenomenon arises because of the high sensitivity of the electronic properties of 2D materials to their atomic arrangement and interlayer interactions. Key factors contributing to this behavior include interlayer coupling, symmetry, and strain. The interlayer coupling significantly alters the band structure, with different stacking types modifying the strength and nature of these interactions, thereby affecting the band gap.\cite{oh2024review} The symmetry of the stacking arrangement influences the overlap of electronic orbitals between layers, where certain configurations enhance or reduce the overlap, leading to a direct or indirect band gap.\cite{vinh2025stacking} Additionally, stacking 2D materials can introduce strain or distortion, which can modify the band structure and potentially convert the band gap between indirect and direct.\cite{vinh2025stacking} Figure~\ref{fig:1} (b, c) also presents the band structures and density of states (DOS) of MoTe$_2$/WSe$_2$ with 0\textdegree~and 60\textdegree~of rotation, respectively. The detailed orbital projection DOS is presented in Figure S14. Both MoTe$_2$ and WSe$_2$ contribute to the band edges at 0\textdegree~of stacking, while MoTe$_2$ and WSe$_2$ dominate the edge of the valence and conduction band respectively at 60\textdegree~of stacking. These findings offer valuable information on the construction of TMDC heterostructures and the tailoring of their properties for electronic applications.

\begin{figure*}
 \centering
 \includegraphics[height=11cm]{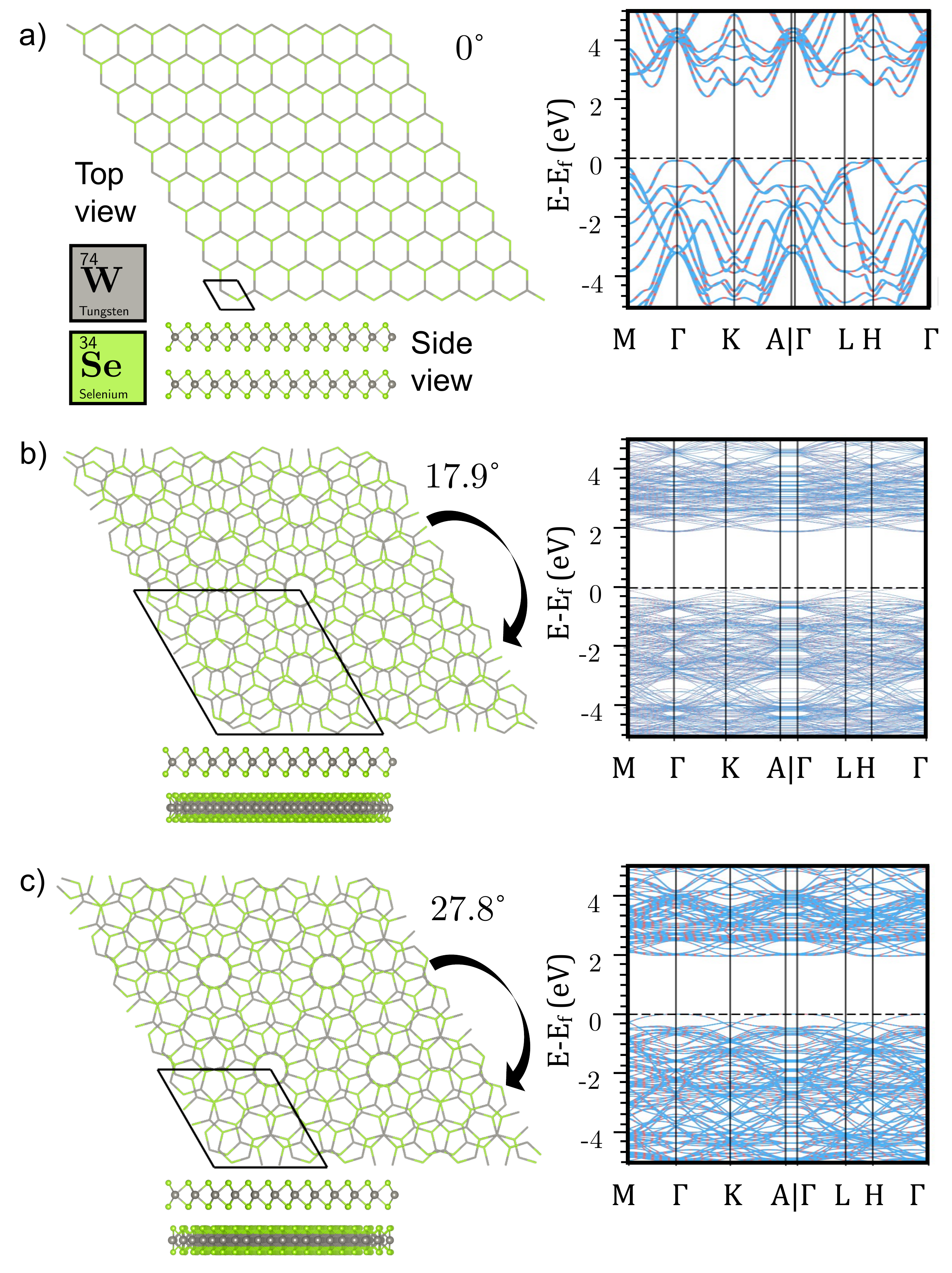}
  \includegraphics[height=11cm]{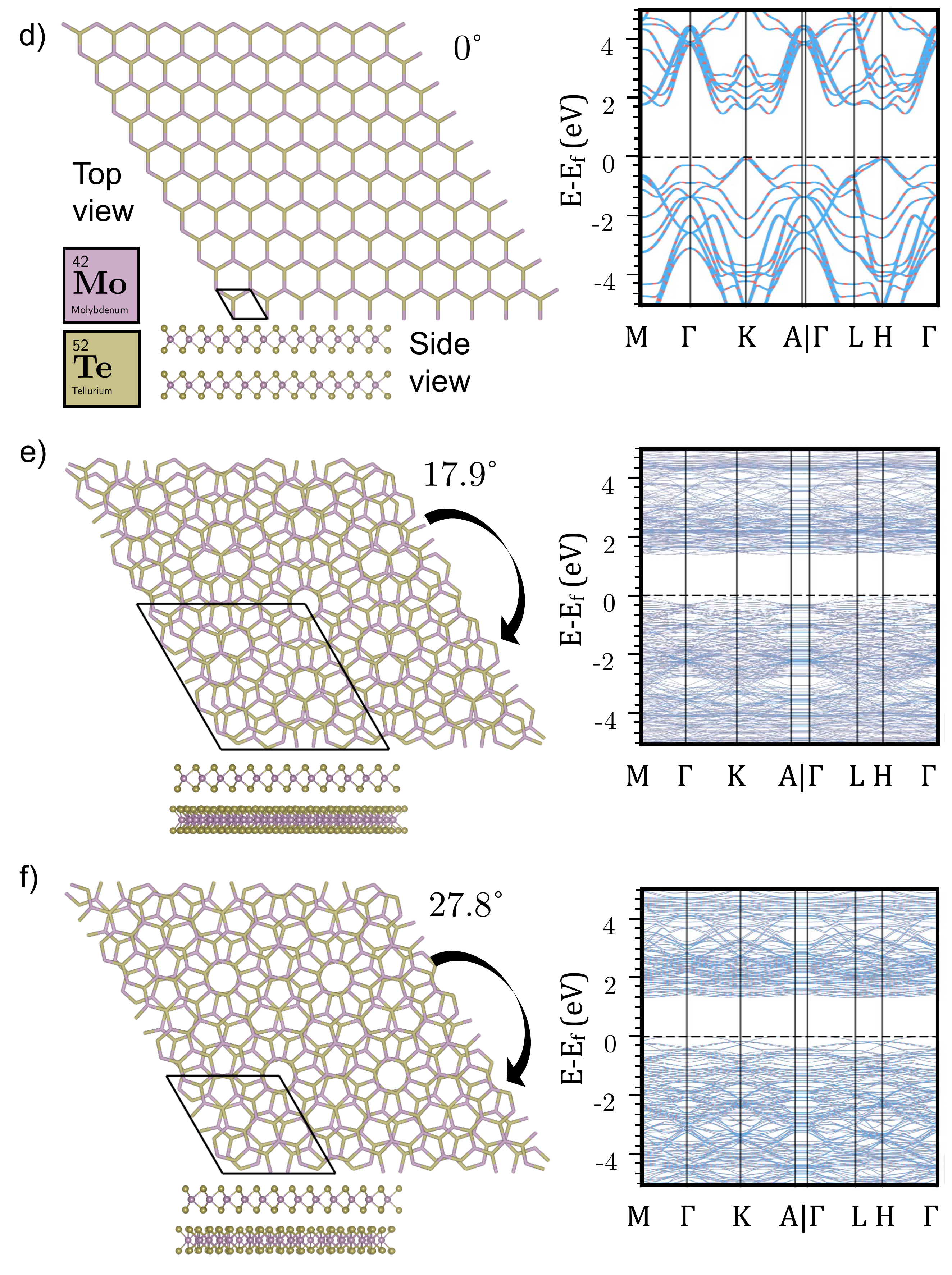}
 \caption{Geometries and band structures of bilayer WSe$_2$ at various twisted angles are shown for (a) 0\textdegree, (b) 17.9\textdegree, (c) 27.8\textdegree as well as \ce{MoTe2} (d-f) for the corresponding angles. Each panel provides both a top view and a side view of the bilayer geometries (shown in the bottom left of each section). For a broader perspective, analogous analyses for other materials including 1H-MoS$_2$, 1T-MoS$_2$, MoSe$_2$, WS$_2$, and WTe$_2$ are comprehensively presented in Figures S7-S12. This extensive representation allows for a detailed comparison across a range of materials and twisted angles, offering a deeper understanding of the relationship between geometric configurations and electronic properties in these layered structures.}
 \label{fig:4}
\end{figure*}

\begin{figure}[h]
\centering
  \includegraphics[width=0.47\textwidth]{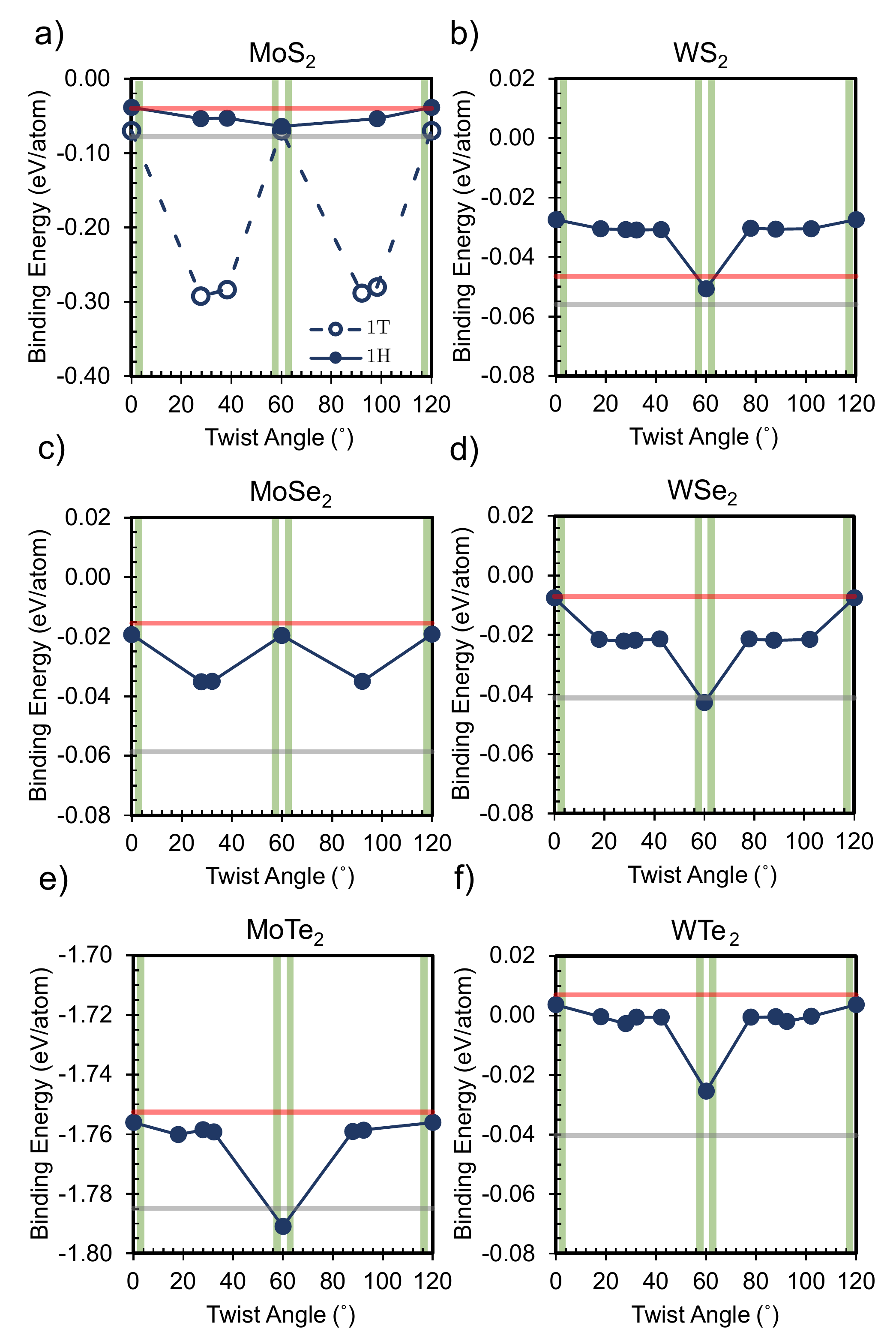}
  \caption{Variation in binding energy as a function of twisted angles for six twisted bilayer cases: (a) MoS$_2$, (b) MoSe$_2$, (c) MoTe$_2$, (d) WS$_2$, (e) WSe$_2$, and (f) WTe$_2$. The red and gray dashed lines denote the binding energies for A-A and A-B stacked homobilayers, respectively, providing a reference for comparison against the fluctuating binding energies observed in the twisted bilayers. The vertical green bars represent the approximate range for the ``Magic Angles'' $\sim 1$ -- $4$~degree. }
  \label{fig:Homo_BE}
\end{figure}

\begin{figure}[h]
\centering
  \includegraphics[width=0.47\textwidth]{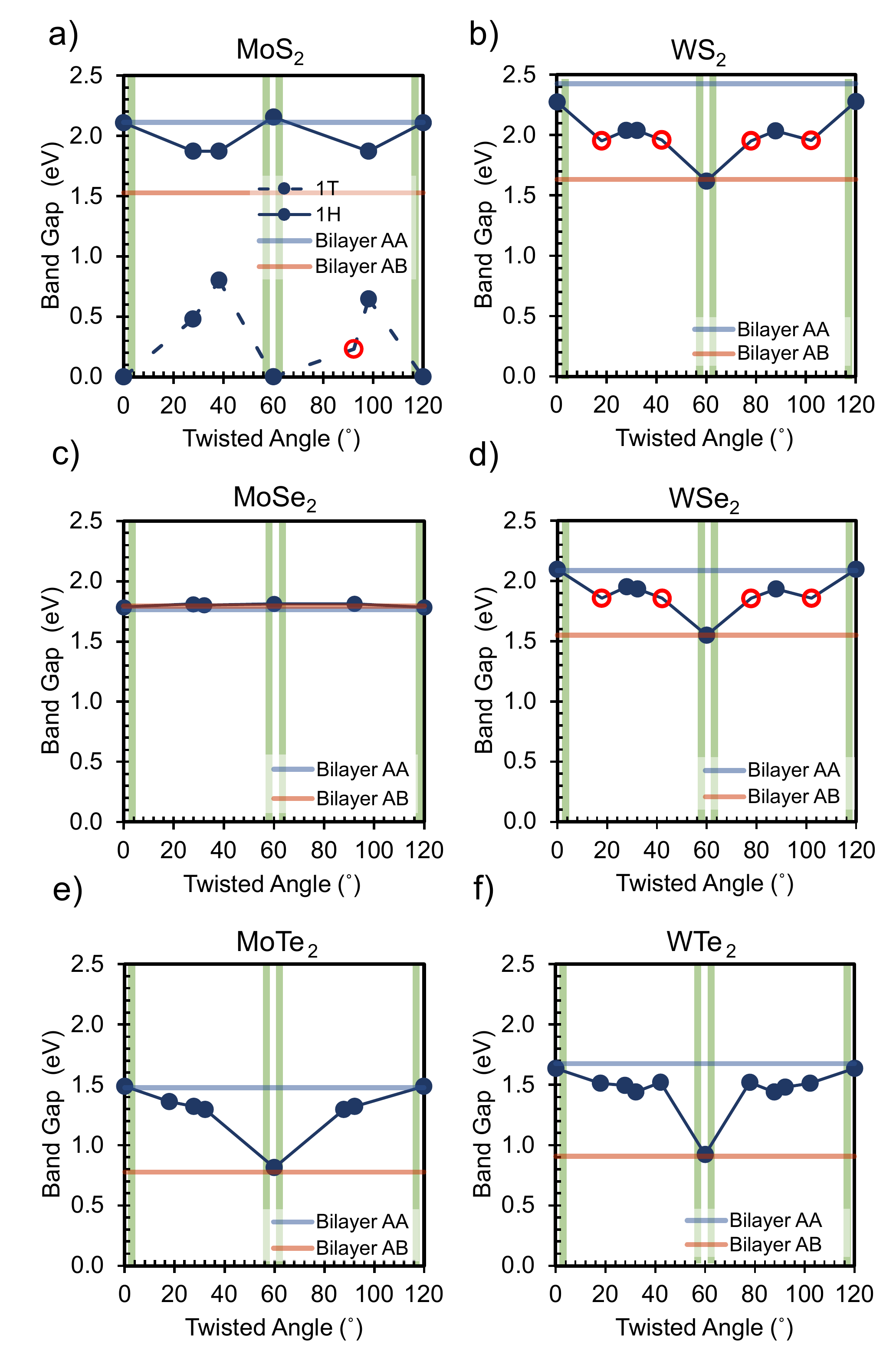}
  \caption{Variation in band gap across a range of twisted angles for six twisted bilayer materials: (a) MoS$_2$, (b) MoSe$_2$, (c) MoTe$_2$, (d) WS$_2$, (e) WSe$_2$, and (f) WTe$_2$. The graph captures the changes in band gap as a function of the twisting angle in these bilayer structures. The vertical green bars again represent the approximate range for the ``Magic Angles''}
  \label{fig:5}
\end{figure}

\subsection{Homo-bilayers}

TMDC heterostructures and homogeneous bilayers offer versatile avenues for engineering material properties. This discussion first addresses fully optimized TMDC bilayers, followed by those with slight strain. We considered six MoX$_2$ TMDCs, each with stacking types A-A and A-B. In A-A stacked bilayers, viewed from above, the cations of each monolayer overlap, whereas in A-B stacking, the cations of one layer overlap with the anions of the other, as depicted in Figure~\ref{fig:3}(a)-(d).
Compared to their monolayer counterparts, the A-A and A-B stacked bilayers of TMDC exhibit unique electronic properties, as summarized in Figure~\ref{fig:3}(e). Key observations include: First, the band gaps of these MX$_2$ TMDCs range from 0.8 to 2.7 eV. The bonding types significantly influence these properties. For example, both monolayer and bilayer forms of 1T-MoS$_2$ consistently exhibit conductive behavior, as shown in Figure S3, a characteristic attributable to its distinctive structure and interlayer coupling, different from 1H-TMDC.
Second, the band gap typically follows the trend: A-B stacking bilayer $<$ A-A stacking bilayer $<$ monolayer.
Third, monolayers of 1H-MoS$_2$, MoSe$_2$, and WS$_2$ present direct band gaps, unlike others.
This evolution from monolayer to bilayer demonstrates significant electronic changes, hinting at the vast potential inherent in multilayered TMDCs for material engineering.
 
In this study, we shifted our focus from conventional A-A and A-B stacking to the more intricate relationships between structure and property in homogeneous TMDC layers, facilitated by varying twisted angles. To refine the scope of our investigation, we initially screened twisted bilayers based on strain magnitude, setting a threshold of less than 0.001 post-formation. The comprehensive trends of strain versus twisted angles for all six TMDCs are depicted in Figure S4. These trends are further elaborated in Figure S5, which zooms in on regions showcasing geometries that meet the strain criteria.
This screening approach is based on the hypothesis of an inverse relationship between structural stability and strain. Subsequently, the low-strain structures underwent symmetrization, and those with more than 40 atoms per unit cell were excluded to optimize computational efficiency. The final candidates selected, adhering to these criteria, are highlighted in Figure S5, marking a strategic narrowing of the potential structures for detailed analysis. This process not only aids in the identification of structurally viable configurations but also sets the stage for a deeper exploration of the nuanced interplay between twisted angles and material properties in TMDC bilayers.

Simultaneously, Moir\'e patterns emerged prominently among these selected candidates. Subsequently, these structures underwent symmetrical optimization while maintaining fixed unit cell sizes, preserving both low strain and the captivating Moir\'e patterns capable of eliciting novel material properties. For example, consider bilayer WSe$_2$: Figure \ref{fig:4} vividly illustrates the evolution of Moir\'e patterns at twisted angles from 0\textdegree to 60\textdegree. Consequently, the band structures exhibit notable variations, with pronounced flattening due to Moir\'e patterns' influence on interlayer coupling interactions. This enhanced flatness facilitates energy conversion by mitigating the impact of crystal momentum. Furthermore, the interlayer interaction of these twisted bilayers was meticulously examined, with Figure~\ref{fig:Homo_BE} showing the binding energy as a function of twisted angles.

In essence, the binding energy of the twisted bilayers exhibits a strong correlation with their geometry, often oscillating between the values observed for A-A and A-B stacked bilayers. Because the A-B-stacked bilayers have a more negative binding energy compared to the A-A-stacked ones, the A-B-stacked interaction is more thermodynamically favorable. A more negative binding energy implies that the system is in a relatively low-energy state, requiring more energy to separate the layers. This phenomenon also occurs in the context of bilayer graphene due to a more favorable interlayer interaction in stacking of A-B\cite{zhou2015van} In particular, except for the WTe$_2$ bilayer with a 0\textdegree~twist, the majority of twisted homobilayers display a negative binding energy within the values of the analogies of stack A-A and A-B. This indicates that most bilayer twistronics in our work can possess a relatively favorable interlayer interaction even compared to conventional stackings. Intriguingly, 1T-MoS$_2$ is predicted to experience a substantial reduction in binding energy at specific twist angles. These findings underscore the role of relatively favorable van der Waals interactions in enabling nonequilibrium stacking configurations, providing insights into the construction of layered twistronics. Furthermore, Figure~\ref{fig:5} visually shows how the band gaps of various TMDC homobilayers varies with twisted angles, revealing different behaviors among different bilayers.

For example, the behavior of 1T-MoS$_2$ is particularly intriguing, as it exhibits a transition between a conductor and a semiconductor depending on the twisted angle, as shown in Figure~\ref{fig:5}(a). Interestingly, at a twist angle of 92.2\textdegree, 1T-MoS$_2$ even achieves a direct band gap with a value of 0.232 eV. In contrast, MoSe$_2$ appears to be less responsive to changes in the band gap-twist relationship, maintaining an indirect bandgap within a limited range of variation (1.79 - 1.81 eV) at different twist angles, as shown in Figure~\ref{fig:5}(b). In contrast, the other four TMDCs (MoTe$_2$, WS$_2$, WSe$_2$, and WTe$_2$) exhibit symmetric band gap-twist relationships, as illustrated in Figure~\ref{fig:5}(c)-(f), with respect to 60\textdegree twist angles. These TMDCs exhibit maximum band gap values at 0\textdegree~ twist and minimum band gap values at 60\textdegree~ twist. In particular, both WS$_2$ and WSe$_2$ undergo a transition from indirect to direct band gap at specific twist angles, namely 17.9\textdegree~ and its corresponding angles (42.1\textdegree~, 77.9\textdegree~and 102.1\textdegree~). These findings highlight that the band gap values of these four twisted TMDCs fluctuate between the band gap values observed in their stacked A-A and A-B bilayers, as demonstrated in the fully optimized bilayer section.

In general, the tunability of these band gaps is evident across a range of twist angles, as:
    1H-MoS$_2$: 1.87 $\sim$ 2.16 eV;
    1T-MoS$_2$: 0.00 $\sim$ 0.80 eV;
    MoSe$_2$: 1.79 $\sim$ 1.81 eV;
    MoTe$_2$: 0.81 $\sim$ 1.49 eV;
    WS$_2$: 1.62 $\sim$ 2.28 eV;
    WSe$_2$: 1.55 $\sim$ 2.10 eV;
    WTe$_2$: 0.92 $\sim$ 1.63 eV. Specifically, 0.2 to 0.3 eV of differences in band gap values can be achieved between a twisting step. Even a difference of 0.1 eV in material band gaps can be significant in various properties and applications.\cite{grundmann2010physics} Not only does a tiny change in band gap affect the conductivity but also influences the optical features. For example, a difference of 0.1 eV can enhance the efficiency of sensors and can change the color of light-emitting diodes (LEDs).\cite{sze2021physics,singh2000semiconductor} Apart from the band gap value, the change in band gap types also has important implications, even if the band gap values are the same. Materials with direct band gaps can efficiently emit and absorb light, whereas materials with indirect band gaps are poor, since the electron transition requires a phonon to conserve momentum.\cite{yuan2018unified} On the other hand, a direct band gap typically has a shorter carrier lifetime because the recombination of electrons and holes can occur more readily. However, the indirect band gap has a longer carrier lifetime, which can be advantageous for applications such as transistors.\cite{yuan2018unified} In short, twisting bilayers offer a sophisticated approach to fine-tuning electronic behavior that can be valuable for different applications. The presence of highly symmetric Moir\'e patterns is believed to induce diverse electronic interactions in low-strain homogeneous bilayer materials.

\section{Methods}
The methodology of our study is divided into two segments: the establishment of prototypes and ab initio calculations. 

\subsection{Establishment of TMDC bilayers}
The heterostructures were generated using \textsc{supercell-forge}, an extension of the \textsc{supercell-core} program\cite{necio2020supercell}, which allows the rapid generation and evaluation of heterostructures at various twist angles. Although the supercell core handles the matching of the in-plane lattice using the Zur and McGill method\cite{Zur1984}, \textsc{supercell-forge} incorporates additional functionality for the optimization of the out-of-plane geometry. The interlayer separation is determined through experimental data-based modeling with the option of applying a machine learning force field for geometry optimization. \textsc{FINDSYM}\cite{stokes2005findsym} is integrated to analyze the underlying symmetries of the generated structures, identifying space groups, lattice parameters and atomic positions within specified tolerances. The program evaluates candidate structures based on multiple criteria, including lattice strain, unit cell size, and total atom count, enabling efficient screening of potential heterostructures. For prototypes of fully relaxed heterostructures and homogeneous bilayers, we confine the rotational angles to either 0\textdegree~ or 60\textdegree, in alignment with the hexagonal symmetry characteristic of the TMDCs studied. We also considered multiple unit cell sizes for each angle, ensuring that the total number of atoms remained below 30 to accommodate computational limitations.

The program \textsc{supercell-forge} was applied to generate the initial models of low-strain yet symmetrized homogeneous bilayers, the candidate structures were selected taking into account structural stress less than 0.001 (Figures S4 and S5) over a range of angles with increments of 0.1\textdegree from 0\textdegree~ to 120\textdegree. These structures were filtered further via the program to fall within certain integer multiples of the lattice parameters, 5, and were restricted to 40 atoms per unit cell, balancing the need for detailed structural representation with computational feasibility. The \textsc{FINDSYM} tolerance parameters within the program were set to a lattice tolerance of 0.001 and 0.01 for the atomic positions. The interlayer distance was determined by a model that included known multilayered TMDC materials. These restrictions allow for materials which are computationally tractable without the use of excessive computational resources and limit the need for considering long-range electron correlation which is outside of the scope of our current DFT-based methods. Finally, the resulting symmetrized materials were allowed to fully relax through the SCF process within \textsc{CRYSTAL23}.

 After such a series of selections, scenarios like twistronics with distorted phases of TMDCs or with small rotational angles are excluded from the targets of this work. These can lead to considerable supercells with low symmetries incalculable by the DFT at the level of hybrid functionals. Examples of such supercells containing around 500 atoms per cell are illustrated in Figure S13. In contrast to studies on bilayer graphene at the magic angle, which employ a combination of DFT at more approximate levels of functionals alongside other approaches, such as tight-binding models or molecular dynamics,\cite{carr2017twistronics,leconte2022relaxation} the DFT approach with hybrid functionals requires massive computational resources for systems containing hundreds of atoms due to the algorithmic scaling of DFT-like methods with the number of orbitals/atoms in the material's asymmetric unit.

\subsection{Computational settings}
In the second phase of our study, the previously established prototypes were subjected to geometric optimization and property analysis using unrestricted and periodic DFT. We used the hybrid functional HSE06\cite{heyd2003hybrid,krukau2006influence}, as implemented in the \textsc{Crystal23} code\cite{dovesi2018quantum}, and integrated it with D3 dispersion correction\cite{grimme2010consistent} and Becke-Jonesson damping\cite{becke2005density}. For Gaussian basis sets, the triple-zeta valence with polarization (TZVP)\cite{vilela2019bsse} was selected. 
During geometric optimization, we performed full optimization for both heterostructures and homogeneous bilayers. For the unconventional homo-bilayers with exotic rotational angles, the geometric optimization was performed with fixed lattice constants. The optimization process adhered to stringent convergence criteria, with an RMS force threshold of 1.54 × 10$^{-2}$ eV/{\AA}, a maximum force of 2.31 × 10$^{-2}$ eV/{\AA}, an RMS displacement of 6.35 × 10$^{-4}$ {\AA}, and a maximum displacement of 9.53 × 10$^{-4}$ {\AA}.
To extract detailed electronic information in the property calculations, we projected the first Brillouin zone ($\sim$ $2\pi$ × $1/60$ {\AA}$^{-1}$) onto a dense Pack-Monkhorst k-mesh grid ($k_a$ × $k_b$ × $k_c$). The mesh grid used much higher shrinking factors (with $a\cdot k_a$ $\ge$ 60, $b\cdot k_b$ $\ge$ 60, $c\cdot k_c$ $\ge$ 60) compared to those used in the geometric optimization phase (where $a\cdot k_a$ $\ge$ 40, $b\cdot k_b$ $\ge$ 40, $c\cdot k_c$ $\ge$ 40).\cite{monkhorst1976special} Also, band structures were computed with the total number of k points along the path set to 1000. These DFT approaches ensured the acquisition of high-resolution data that is essential for a comprehensive understanding of the electronic properties of these complex systems. Apart from the DFT algorithm settings, we executed the computation with 32 processors (Intel Xeon Gold 6148 Processor) on one node. The CPU time requires a geometric optimization ranging from 2 to 16 million seconds.

The calculation of the binding energy per atom ($E_{BE}$) for the bilayers in our study is performed using the following equation:
\begin{equation}\label{eqn:cohesive}
    E_{BE} = 2\frac{E_{BL}}{N_{BL}} - \frac{E_{ML 1}}{N_{ML 1}} - \frac{E_{ML 2}}{N_{ML 2}}
\end{equation}
In this equation, $E_{BL}$ denotes the total energy per unit cell of the bilayer, while $E_{ML 1}$ and $E_{ML 2}$ represent the total energies per unit cell of the first and second monolayer building blocks, respectively. The terms $N_{BL}$, $N_{ML 1}$, and $N_{ML 2}$ correspond to the number of atoms in the unit cell of the bilayer and the first and second monolayers, respectively.

\section{Conclusion}

In summary, this work demonstrates how manipulating the stacking and twist angles in bilayer TMDCs (MX$_2$, with M = Mo, W and X = S, Se, Te) unlocks a broad spectrum of electronic and thermodynamic behaviors. By combining systematic screening of low-strain supercells and range-separated hybrid DFT calculations, we map out how weak van der Waals interactions can be exploited to fine-tune the band gap magnitudes, conductivity, and direct--indirect gap transitions across both heterobilayers and homobilayers. In particular, the MoTe$_2$/WSe$_2$ heterobilayer undergoes a striking transformation from indirect to direct band gaps, underscoring its potential in energy-conversion devices. Meanwhile, 1T-MoS$_2$ exhibits a unique conductor--semiconductor switching at certain twist angles, revealing intriguing prospects for electronic switching applications.

The emergence of Moir\'e patterns at specific twist angles plays a central role in flattening the band structure, thus enhancing carrier interactions and, in some cases, facilitating the design of more efficient optical and electronic devices. Early experiments \cite{Regan2020-po,Tang2020-pm} in twisted TMDC systems provided key insights into how smaller Moir\'e patterns can induce band structure modifications without immediate topological features. More recent developments indicate that larger periodicities can lead to the formation of fractional Chern insulators and correlated states in twisted bilayer MoTe$_2$ \cite{yu2024fractional,jia2024moire}. By systematically varying angles up to 60\textdegree, we identify critical twists---e.g., around 17.9\textdegree~in WS$_2$ and WSe$_2$ ---that induce direct band gaps, expanding the design possibilities for light emission and sensing technologies. Moreover, our binding energy analysis confirms that MoTe$_2$-based bilayers exhibit robust interlayer coupling, while WTe$_2$-based bilayers are relatively less stable, demonstrating the intricate interplay between material composition and thermodynamic favorability.

In the future, to fully realize the potential of twisted TMDC bilayers, several critical directions must be pursued. Spin-orbit coupling effects must be explicitly included, as they significantly influence the band topology and electronic properties of these systems. Additionally, the limitations of range-separated hybrid functionals like HSE06, particularly the self-interaction and Koopmans' condition errors \cite{Holmes2015-jb}, necessitate the exploration of higher-level functionals and beyond-DFT methods, including exact diagonalization \cite{medvedeva2017exact,Lu2017-ts}, embedded dynamical mean-field theory (DMFT) \cite{Kotliar2004-pr}, and quantum Monte Carlo approaches \cite{haule2007quantum,Bauer2011-ed}. Developing a robust band unfolding procedure for \textsc{Crystal23} is crucial to examine large Moir\'e lattices, allowing a detailed analysis of electronic properties. Furthermore, rigorous benchmarking of DFT-based methods for twisted TMDCs with varying Moir\'e lattice parameters will provide a reliable framework for evaluating and predicting their properties.

By illuminating how stacking configurations, twist angles, and material choices converge to shape electronic properties, this study sets the stage for designing next-generation 2D materials. Whether for energy-harvesting, optoelectronics, or quantum applications, twisted TMDC bilayers represent a versatile and rapidly evolving platform poised to reveal new phases of matter and to enable a range of transformative device technologies.

\bibliography{main}

\begin{thebibliography}{64}%
\makeatletter
\providecommand \@ifxundefined [1]{%
 \@ifx{#1\undefined}
}%
\providecommand \@ifnum [1]{%
 \ifnum #1\expandafter \@firstoftwo
 \else \expandafter \@secondoftwo
 \fi
}%
\providecommand \@ifx [1]{%
 \ifx #1\expandafter \@firstoftwo
 \else \expandafter \@secondoftwo
 \fi
}%
\providecommand \natexlab [1]{#1}%
\providecommand \enquote  [1]{``#1''}%
\providecommand \bibnamefont  [1]{#1}%
\providecommand \bibfnamefont [1]{#1}%
\providecommand \citenamefont [1]{#1}%
\providecommand \href@noop [0]{\@secondoftwo}%
\providecommand \href [0]{\begingroup \@sanitize@url \@href}%
\providecommand \@href[1]{\@@startlink{#1}\@@href}%
\providecommand \@@href[1]{\endgroup#1\@@endlink}%
\providecommand \@sanitize@url [0]{\catcode `\\12\catcode `\$12\catcode
  `\&12\catcode `\#12\catcode `\^12\catcode `\_12\catcode `\%12\relax}%
\providecommand \@@startlink[1]{}%
\providecommand \@@endlink[0]{}%
\providecommand \url  [0]{\begingroup\@sanitize@url \@url }%
\providecommand \@url [1]{\endgroup\@href {#1}{\urlprefix }}%
\providecommand \urlprefix  [0]{URL }%
\providecommand \Eprint [0]{\href }%
\providecommand \doibase [0]{https://doi.org/}%
\providecommand \selectlanguage [0]{\@gobble}%
\providecommand \bibinfo  [0]{\@secondoftwo}%
\providecommand \bibfield  [0]{\@secondoftwo}%
\providecommand \translation [1]{[#1]}%
\providecommand \BibitemOpen [0]{}%
\providecommand \bibitemStop [0]{}%
\providecommand \bibitemNoStop [0]{.\EOS\space}%
\providecommand \EOS [0]{\spacefactor3000\relax}%
\providecommand \BibitemShut  [1]{\csname bibitem#1\endcsname}%
\let\auto@bib@innerbib\@empty
\bibitem [{\citenamefont {Voiry}\ \emph {et~al.}(2015)\citenamefont {Voiry},
  \citenamefont {Mohite},\ and\ \citenamefont {Chhowalla}}]{voiry2015phase}%
  \BibitemOpen
  \bibfield  {author} {\bibinfo {author} {\bibfnamefont {D.}~\bibnamefont
  {Voiry}}, \bibinfo {author} {\bibfnamefont {A.}~\bibnamefont {Mohite}},\ and\
  \bibinfo {author} {\bibfnamefont {M.}~\bibnamefont {Chhowalla}},\ }\href@noop
  {} {\bibfield  {journal} {\bibinfo  {journal} {Chemical Society Reviews}\
  }\textbf {\bibinfo {volume} {44}},\ \bibinfo {pages} {2702} (\bibinfo {year}
  {2015})}\BibitemShut {NoStop}%
\bibitem [{\citenamefont {Long}\ \emph {et~al.}(2019)\citenamefont {Long},
  \citenamefont {Dai}, \citenamefont {Gong},\ and\ \citenamefont
  {Jin}}]{long2019robust}%
  \BibitemOpen
  \bibfield  {author} {\bibinfo {author} {\bibfnamefont {C.}~\bibnamefont
  {Long}}, \bibinfo {author} {\bibfnamefont {Y.}~\bibnamefont {Dai}}, \bibinfo
  {author} {\bibfnamefont {Z.-R.}\ \bibnamefont {Gong}},\ and\ \bibinfo
  {author} {\bibfnamefont {H.}~\bibnamefont {Jin}},\ }\href@noop {} {\bibfield
  {journal} {\bibinfo  {journal} {Physical Review B}\ }\textbf {\bibinfo
  {volume} {99}},\ \bibinfo {pages} {115316} (\bibinfo {year}
  {2019})}\BibitemShut {NoStop}%
\bibitem [{\citenamefont {Shi}\ \emph {et~al.}(2018)\citenamefont {Shi},
  \citenamefont {Wang}, \citenamefont {Sun}, \citenamefont {Li},\ and\
  \citenamefont {Zhang}}]{shi2018interlayer}%
  \BibitemOpen
  \bibfield  {author} {\bibinfo {author} {\bibfnamefont {Z.}~\bibnamefont
  {Shi}}, \bibinfo {author} {\bibfnamefont {X.}~\bibnamefont {Wang}}, \bibinfo
  {author} {\bibfnamefont {Y.}~\bibnamefont {Sun}}, \bibinfo {author}
  {\bibfnamefont {Y.}~\bibnamefont {Li}},\ and\ \bibinfo {author}
  {\bibfnamefont {L.}~\bibnamefont {Zhang}},\ }\href@noop {} {\bibfield
  {journal} {\bibinfo  {journal} {Semiconductor Science and Technology}\
  }\textbf {\bibinfo {volume} {33}},\ \bibinfo {pages} {093001} (\bibinfo
  {year} {2018})}\BibitemShut {NoStop}%
\bibitem [{\citenamefont {Dean}\ \emph {et~al.}(2010)\citenamefont {Dean},
  \citenamefont {Young}, \citenamefont {Meric}, \citenamefont {Lee},
  \citenamefont {Wang}, \citenamefont {Sorgenfrei}, \citenamefont {Watanabe},
  \citenamefont {Taniguchi}, \citenamefont {Kim}, \citenamefont {Shepard} \emph
  {et~al.}}]{dean2010boron}%
  \BibitemOpen
  \bibfield  {author} {\bibinfo {author} {\bibfnamefont {C.~R.}\ \bibnamefont
  {Dean}}, \bibinfo {author} {\bibfnamefont {A.~F.}\ \bibnamefont {Young}},
  \bibinfo {author} {\bibfnamefont {I.}~\bibnamefont {Meric}}, \bibinfo
  {author} {\bibfnamefont {C.}~\bibnamefont {Lee}}, \bibinfo {author}
  {\bibfnamefont {L.}~\bibnamefont {Wang}}, \bibinfo {author} {\bibfnamefont
  {S.}~\bibnamefont {Sorgenfrei}}, \bibinfo {author} {\bibfnamefont
  {K.}~\bibnamefont {Watanabe}}, \bibinfo {author} {\bibfnamefont
  {T.}~\bibnamefont {Taniguchi}}, \bibinfo {author} {\bibfnamefont
  {P.}~\bibnamefont {Kim}}, \bibinfo {author} {\bibfnamefont {K.~L.}\
  \bibnamefont {Shepard}}, \emph {et~al.},\ }\href@noop {} {\bibfield
  {journal} {\bibinfo  {journal} {Nature nanotechnology}\ }\textbf {\bibinfo
  {volume} {5}},\ \bibinfo {pages} {722} (\bibinfo {year} {2010})}\BibitemShut
  {NoStop}%
\bibitem [{\citenamefont {Sahoo}\ \emph {et~al.}(2018)\citenamefont {Sahoo},
  \citenamefont {Memaran}, \citenamefont {Xin}, \citenamefont {Balicas},\ and\
  \citenamefont {Guti{\'e}rrez}}]{sahoo2018one}%
  \BibitemOpen
  \bibfield  {author} {\bibinfo {author} {\bibfnamefont {P.~K.}\ \bibnamefont
  {Sahoo}}, \bibinfo {author} {\bibfnamefont {S.}~\bibnamefont {Memaran}},
  \bibinfo {author} {\bibfnamefont {Y.}~\bibnamefont {Xin}}, \bibinfo {author}
  {\bibfnamefont {L.}~\bibnamefont {Balicas}},\ and\ \bibinfo {author}
  {\bibfnamefont {H.~R.}\ \bibnamefont {Guti{\'e}rrez}},\ }\href@noop {}
  {\bibfield  {journal} {\bibinfo  {journal} {Nature}\ }\textbf {\bibinfo
  {volume} {553}},\ \bibinfo {pages} {63} (\bibinfo {year} {2018})}\BibitemShut
  {NoStop}%
\bibitem [{\citenamefont {Jin}\ \emph {et~al.}(2019)\citenamefont {Jin},
  \citenamefont {Regan}, \citenamefont {Yan}, \citenamefont {Iqbal
  Bakti~Utama}, \citenamefont {Wang}, \citenamefont {Zhao}, \citenamefont
  {Qin}, \citenamefont {Yang}, \citenamefont {Zheng}, \citenamefont {Shi} \emph
  {et~al.}}]{jin2019observation}%
  \BibitemOpen
  \bibfield  {author} {\bibinfo {author} {\bibfnamefont {C.}~\bibnamefont
  {Jin}}, \bibinfo {author} {\bibfnamefont {E.~C.}\ \bibnamefont {Regan}},
  \bibinfo {author} {\bibfnamefont {A.}~\bibnamefont {Yan}}, \bibinfo {author}
  {\bibfnamefont {M.}~\bibnamefont {Iqbal Bakti~Utama}}, \bibinfo {author}
  {\bibfnamefont {D.}~\bibnamefont {Wang}}, \bibinfo {author} {\bibfnamefont
  {S.}~\bibnamefont {Zhao}}, \bibinfo {author} {\bibfnamefont {Y.}~\bibnamefont
  {Qin}}, \bibinfo {author} {\bibfnamefont {S.}~\bibnamefont {Yang}}, \bibinfo
  {author} {\bibfnamefont {Z.}~\bibnamefont {Zheng}}, \bibinfo {author}
  {\bibfnamefont {S.}~\bibnamefont {Shi}}, \emph {et~al.},\ }\href@noop {}
  {\bibfield  {journal} {\bibinfo  {journal} {Nature}\ }\textbf {\bibinfo
  {volume} {567}},\ \bibinfo {pages} {76} (\bibinfo {year} {2019})}\BibitemShut
  {NoStop}%
\bibitem [{\citenamefont {Terrones}\ and\ \citenamefont
  {Terrones}(2014)}]{terrones2014bilayers}%
  \BibitemOpen
  \bibfield  {author} {\bibinfo {author} {\bibfnamefont {H.}~\bibnamefont
  {Terrones}}\ and\ \bibinfo {author} {\bibfnamefont {M.}~\bibnamefont
  {Terrones}},\ }\href@noop {} {\bibfield  {journal} {\bibinfo  {journal}
  {Journal of Materials Research}\ }\textbf {\bibinfo {volume} {29}},\ \bibinfo
  {pages} {373} (\bibinfo {year} {2014})}\BibitemShut {NoStop}%
\bibitem [{\citenamefont {Wang}\ \emph {et~al.}(2016)\citenamefont {Wang},
  \citenamefont {Huang}, \citenamefont {Tian}, \citenamefont {Ceballos},
  \citenamefont {Lin}, \citenamefont {Mahjouri-Samani}, \citenamefont
  {Boulesbaa}, \citenamefont {Puretzky}, \citenamefont {Rouleau}, \citenamefont
  {Yoon} \emph {et~al.}}]{wang2016interlayer}%
  \BibitemOpen
  \bibfield  {author} {\bibinfo {author} {\bibfnamefont {K.}~\bibnamefont
  {Wang}}, \bibinfo {author} {\bibfnamefont {B.}~\bibnamefont {Huang}},
  \bibinfo {author} {\bibfnamefont {M.}~\bibnamefont {Tian}}, \bibinfo {author}
  {\bibfnamefont {F.}~\bibnamefont {Ceballos}}, \bibinfo {author}
  {\bibfnamefont {M.-W.}\ \bibnamefont {Lin}}, \bibinfo {author} {\bibfnamefont
  {M.}~\bibnamefont {Mahjouri-Samani}}, \bibinfo {author} {\bibfnamefont
  {A.}~\bibnamefont {Boulesbaa}}, \bibinfo {author} {\bibfnamefont {A.~A.}\
  \bibnamefont {Puretzky}}, \bibinfo {author} {\bibfnamefont {C.~M.}\
  \bibnamefont {Rouleau}}, \bibinfo {author} {\bibfnamefont {M.}~\bibnamefont
  {Yoon}}, \emph {et~al.},\ }\href@noop {} {\bibfield  {journal} {\bibinfo
  {journal} {ACS nano}\ }\textbf {\bibinfo {volume} {10}},\ \bibinfo {pages}
  {6612} (\bibinfo {year} {2016})}\BibitemShut {NoStop}%
\bibitem [{\citenamefont {Jones}\ \emph {et~al.}(2014)\citenamefont {Jones},
  \citenamefont {Yu}, \citenamefont {Ross}, \citenamefont {Klement},
  \citenamefont {Ghimire}, \citenamefont {Yan}, \citenamefont {Mandrus},
  \citenamefont {Yao},\ and\ \citenamefont {Xu}}]{jones2014spin}%
  \BibitemOpen
  \bibfield  {author} {\bibinfo {author} {\bibfnamefont {A.~M.}\ \bibnamefont
  {Jones}}, \bibinfo {author} {\bibfnamefont {H.}~\bibnamefont {Yu}}, \bibinfo
  {author} {\bibfnamefont {J.~S.}\ \bibnamefont {Ross}}, \bibinfo {author}
  {\bibfnamefont {P.}~\bibnamefont {Klement}}, \bibinfo {author} {\bibfnamefont
  {N.~J.}\ \bibnamefont {Ghimire}}, \bibinfo {author} {\bibfnamefont
  {J.}~\bibnamefont {Yan}}, \bibinfo {author} {\bibfnamefont {D.~G.}\
  \bibnamefont {Mandrus}}, \bibinfo {author} {\bibfnamefont {W.}~\bibnamefont
  {Yao}},\ and\ \bibinfo {author} {\bibfnamefont {X.}~\bibnamefont {Xu}},\
  }\href@noop {} {\bibfield  {journal} {\bibinfo  {journal} {Nature Physics}\
  }\textbf {\bibinfo {volume} {10}},\ \bibinfo {pages} {130} (\bibinfo {year}
  {2014})}\BibitemShut {NoStop}%
\bibitem [{\citenamefont {Sung}\ \emph {et~al.}(2020)\citenamefont {Sung},
  \citenamefont {Zhou}, \citenamefont {Scuri}, \citenamefont {Z{\'o}lyomi},
  \citenamefont {Andersen}, \citenamefont {Yoo}, \citenamefont {Wild},
  \citenamefont {Joe}, \citenamefont {Gelly}, \citenamefont {Heo} \emph
  {et~al.}}]{sung2020broken}%
  \BibitemOpen
  \bibfield  {author} {\bibinfo {author} {\bibfnamefont {J.}~\bibnamefont
  {Sung}}, \bibinfo {author} {\bibfnamefont {Y.}~\bibnamefont {Zhou}}, \bibinfo
  {author} {\bibfnamefont {G.}~\bibnamefont {Scuri}}, \bibinfo {author}
  {\bibfnamefont {V.}~\bibnamefont {Z{\'o}lyomi}}, \bibinfo {author}
  {\bibfnamefont {T.~I.}\ \bibnamefont {Andersen}}, \bibinfo {author}
  {\bibfnamefont {H.}~\bibnamefont {Yoo}}, \bibinfo {author} {\bibfnamefont
  {D.~S.}\ \bibnamefont {Wild}}, \bibinfo {author} {\bibfnamefont {A.~Y.}\
  \bibnamefont {Joe}}, \bibinfo {author} {\bibfnamefont {R.~J.}\ \bibnamefont
  {Gelly}}, \bibinfo {author} {\bibfnamefont {H.}~\bibnamefont {Heo}}, \emph
  {et~al.},\ }\href@noop {} {\bibfield  {journal} {\bibinfo  {journal} {Nature
  Nanotechnology}\ }\textbf {\bibinfo {volume} {15}},\ \bibinfo {pages} {750}
  (\bibinfo {year} {2020})}\BibitemShut {NoStop}%
\bibitem [{\citenamefont {Xiao}\ \emph {et~al.}(2012)\citenamefont {Xiao},
  \citenamefont {Liu}, \citenamefont {Feng}, \citenamefont {Xu},\ and\
  \citenamefont {Yao}}]{xiao2012coupled}%
  \BibitemOpen
  \bibfield  {author} {\bibinfo {author} {\bibfnamefont {D.}~\bibnamefont
  {Xiao}}, \bibinfo {author} {\bibfnamefont {G.-B.}\ \bibnamefont {Liu}},
  \bibinfo {author} {\bibfnamefont {W.}~\bibnamefont {Feng}}, \bibinfo {author}
  {\bibfnamefont {X.}~\bibnamefont {Xu}},\ and\ \bibinfo {author}
  {\bibfnamefont {W.}~\bibnamefont {Yao}},\ }\href@noop {} {\bibfield
  {journal} {\bibinfo  {journal} {Physical review letters}\ }\textbf {\bibinfo
  {volume} {108}},\ \bibinfo {pages} {196802} (\bibinfo {year}
  {2012})}\BibitemShut {NoStop}%
\bibitem [{\citenamefont {Li}\ and\ \citenamefont {Wu}(2017)}]{li2017binary}%
  \BibitemOpen
  \bibfield  {author} {\bibinfo {author} {\bibfnamefont {L.}~\bibnamefont
  {Li}}\ and\ \bibinfo {author} {\bibfnamefont {M.}~\bibnamefont {Wu}},\
  }\href@noop {} {\bibfield  {journal} {\bibinfo  {journal} {ACS nano}\
  }\textbf {\bibinfo {volume} {11}},\ \bibinfo {pages} {6382} (\bibinfo {year}
  {2017})}\BibitemShut {NoStop}%
\bibitem [{\citenamefont {Zhang}\ \emph {et~al.}(2019)\citenamefont {Zhang},
  \citenamefont {Nan}, \citenamefont {Xiao}, \citenamefont {Wan}, \citenamefont
  {Gu}, \citenamefont {Du}, \citenamefont {Ni},\ and\ \citenamefont
  {Ostrikov}}]{zhang2019transition}%
  \BibitemOpen
  \bibfield  {author} {\bibinfo {author} {\bibfnamefont {X.}~\bibnamefont
  {Zhang}}, \bibinfo {author} {\bibfnamefont {H.}~\bibnamefont {Nan}}, \bibinfo
  {author} {\bibfnamefont {S.}~\bibnamefont {Xiao}}, \bibinfo {author}
  {\bibfnamefont {X.}~\bibnamefont {Wan}}, \bibinfo {author} {\bibfnamefont
  {X.}~\bibnamefont {Gu}}, \bibinfo {author} {\bibfnamefont {A.}~\bibnamefont
  {Du}}, \bibinfo {author} {\bibfnamefont {Z.}~\bibnamefont {Ni}},\ and\
  \bibinfo {author} {\bibfnamefont {K.}~\bibnamefont {Ostrikov}},\ }\href@noop
  {} {\bibfield  {journal} {\bibinfo  {journal} {Nature communications}\
  }\textbf {\bibinfo {volume} {10}},\ \bibinfo {pages} {598} (\bibinfo {year}
  {2019})}\BibitemShut {NoStop}%
\bibitem [{\citenamefont {Kim}\ \emph {et~al.}(2012)\citenamefont {Kim},
  \citenamefont {Konar}, \citenamefont {Hwang}, \citenamefont {Lee},
  \citenamefont {Lee}, \citenamefont {Yang}, \citenamefont {Jung},
  \citenamefont {Kim}, \citenamefont {Yoo}, \citenamefont {Choi} \emph
  {et~al.}}]{kim2012high}%
  \BibitemOpen
  \bibfield  {author} {\bibinfo {author} {\bibfnamefont {S.}~\bibnamefont
  {Kim}}, \bibinfo {author} {\bibfnamefont {A.}~\bibnamefont {Konar}}, \bibinfo
  {author} {\bibfnamefont {W.-S.}\ \bibnamefont {Hwang}}, \bibinfo {author}
  {\bibfnamefont {J.~H.}\ \bibnamefont {Lee}}, \bibinfo {author} {\bibfnamefont
  {J.}~\bibnamefont {Lee}}, \bibinfo {author} {\bibfnamefont {J.}~\bibnamefont
  {Yang}}, \bibinfo {author} {\bibfnamefont {C.}~\bibnamefont {Jung}}, \bibinfo
  {author} {\bibfnamefont {H.}~\bibnamefont {Kim}}, \bibinfo {author}
  {\bibfnamefont {J.-B.}\ \bibnamefont {Yoo}}, \bibinfo {author} {\bibfnamefont
  {J.-Y.}\ \bibnamefont {Choi}}, \emph {et~al.},\ }\href@noop {} {\bibfield
  {journal} {\bibinfo  {journal} {Nature communications}\ }\textbf {\bibinfo
  {volume} {3}},\ \bibinfo {pages} {1011} (\bibinfo {year} {2012})}\BibitemShut
  {NoStop}%
\bibitem [{\citenamefont {Tang}\ \emph {et~al.}(2021)\citenamefont {Tang},
  \citenamefont {Che}, \citenamefont {Xu}, \citenamefont {Ang}, \citenamefont
  {Di}, \citenamefont {Gao}, \citenamefont {Yang}, \citenamefont {Zhou},\ and\
  \citenamefont {Liu}}]{tang2021recent}%
  \BibitemOpen
  \bibfield  {author} {\bibinfo {author} {\bibfnamefont {B.}~\bibnamefont
  {Tang}}, \bibinfo {author} {\bibfnamefont {B.}~\bibnamefont {Che}}, \bibinfo
  {author} {\bibfnamefont {M.}~\bibnamefont {Xu}}, \bibinfo {author}
  {\bibfnamefont {Z.~P.}\ \bibnamefont {Ang}}, \bibinfo {author} {\bibfnamefont
  {J.}~\bibnamefont {Di}}, \bibinfo {author} {\bibfnamefont {H.-J.}\
  \bibnamefont {Gao}}, \bibinfo {author} {\bibfnamefont {H.}~\bibnamefont
  {Yang}}, \bibinfo {author} {\bibfnamefont {J.}~\bibnamefont {Zhou}},\ and\
  \bibinfo {author} {\bibfnamefont {Z.}~\bibnamefont {Liu}},\ }\href@noop {}
  {\bibfield  {journal} {\bibinfo  {journal} {Small Structures}\ }\textbf
  {\bibinfo {volume} {2}},\ \bibinfo {pages} {2000153} (\bibinfo {year}
  {2021})}\BibitemShut {NoStop}%
\bibitem [{\citenamefont {Lin}\ \emph {et~al.}(2023)\citenamefont {Lin},
  \citenamefont {Torsi}, \citenamefont {Younas}, \citenamefont {Hinkle},
  \citenamefont {Rigosi}, \citenamefont {Hill}, \citenamefont {Zhang},
  \citenamefont {Huang}, \citenamefont {Shuck}, \citenamefont {Chen} \emph
  {et~al.}}]{lin2023recent}%
  \BibitemOpen
  \bibfield  {author} {\bibinfo {author} {\bibfnamefont {Y.-C.}\ \bibnamefont
  {Lin}}, \bibinfo {author} {\bibfnamefont {R.}~\bibnamefont {Torsi}}, \bibinfo
  {author} {\bibfnamefont {R.}~\bibnamefont {Younas}}, \bibinfo {author}
  {\bibfnamefont {C.~L.}\ \bibnamefont {Hinkle}}, \bibinfo {author}
  {\bibfnamefont {A.~F.}\ \bibnamefont {Rigosi}}, \bibinfo {author}
  {\bibfnamefont {H.~M.}\ \bibnamefont {Hill}}, \bibinfo {author}
  {\bibfnamefont {K.}~\bibnamefont {Zhang}}, \bibinfo {author} {\bibfnamefont
  {S.}~\bibnamefont {Huang}}, \bibinfo {author} {\bibfnamefont {C.~E.}\
  \bibnamefont {Shuck}}, \bibinfo {author} {\bibfnamefont {C.}~\bibnamefont
  {Chen}}, \emph {et~al.},\ }\href@noop {} {\bibfield  {journal} {\bibinfo
  {journal} {ACS nano}\ }\textbf {\bibinfo {volume} {17}},\ \bibinfo {pages}
  {9694} (\bibinfo {year} {2023})}\BibitemShut {NoStop}%
\bibitem [{\citenamefont {Cao}\ \emph {et~al.}(2018{\natexlab{a}})\citenamefont
  {Cao}, \citenamefont {Fatemi}, \citenamefont {Fang}, \citenamefont
  {Watanabe}, \citenamefont {Taniguchi}, \citenamefont {Kaxiras},\ and\
  \citenamefont {Jarillo-Herrero}}]{cao2018unconventional}%
  \BibitemOpen
  \bibfield  {author} {\bibinfo {author} {\bibfnamefont {Y.}~\bibnamefont
  {Cao}}, \bibinfo {author} {\bibfnamefont {V.}~\bibnamefont {Fatemi}},
  \bibinfo {author} {\bibfnamefont {S.}~\bibnamefont {Fang}}, \bibinfo {author}
  {\bibfnamefont {K.}~\bibnamefont {Watanabe}}, \bibinfo {author}
  {\bibfnamefont {T.}~\bibnamefont {Taniguchi}}, \bibinfo {author}
  {\bibfnamefont {E.}~\bibnamefont {Kaxiras}},\ and\ \bibinfo {author}
  {\bibfnamefont {P.}~\bibnamefont {Jarillo-Herrero}},\ }\href@noop {}
  {\bibfield  {journal} {\bibinfo  {journal} {Nature}\ }\textbf {\bibinfo
  {volume} {556}},\ \bibinfo {pages} {43} (\bibinfo {year}
  {2018}{\natexlab{a}})}\BibitemShut {NoStop}%
\bibitem [{\citenamefont {Cao}\ \emph {et~al.}(2018{\natexlab{b}})\citenamefont
  {Cao}, \citenamefont {Fatemi}, \citenamefont {Demir}, \citenamefont {Fang},
  \citenamefont {Tomarken}, \citenamefont {Luo}, \citenamefont
  {Sanchez-Yamagishi}, \citenamefont {Watanabe}, \citenamefont {Taniguchi},
  \citenamefont {Kaxiras} \emph {et~al.}}]{cao2018correlated}%
  \BibitemOpen
  \bibfield  {author} {\bibinfo {author} {\bibfnamefont {Y.}~\bibnamefont
  {Cao}}, \bibinfo {author} {\bibfnamefont {V.}~\bibnamefont {Fatemi}},
  \bibinfo {author} {\bibfnamefont {A.}~\bibnamefont {Demir}}, \bibinfo
  {author} {\bibfnamefont {S.}~\bibnamefont {Fang}}, \bibinfo {author}
  {\bibfnamefont {S.~L.}\ \bibnamefont {Tomarken}}, \bibinfo {author}
  {\bibfnamefont {J.~Y.}\ \bibnamefont {Luo}}, \bibinfo {author} {\bibfnamefont
  {J.~D.}\ \bibnamefont {Sanchez-Yamagishi}}, \bibinfo {author} {\bibfnamefont
  {K.}~\bibnamefont {Watanabe}}, \bibinfo {author} {\bibfnamefont
  {T.}~\bibnamefont {Taniguchi}}, \bibinfo {author} {\bibfnamefont
  {E.}~\bibnamefont {Kaxiras}}, \emph {et~al.},\ }\href@noop {} {\bibfield
  {journal} {\bibinfo  {journal} {Nature}\ }\textbf {\bibinfo {volume} {556}},\
  \bibinfo {pages} {80} (\bibinfo {year} {2018}{\natexlab{b}})}\BibitemShut
  {NoStop}%
\bibitem [{\citenamefont {Wu}\ \emph {et~al.}(2019)\citenamefont {Wu},
  \citenamefont {Lovorn}, \citenamefont {Tutuc}, \citenamefont {Martin},\ and\
  \citenamefont {MacDonald}}]{wu2019topological}%
  \BibitemOpen
  \bibfield  {author} {\bibinfo {author} {\bibfnamefont {F.}~\bibnamefont
  {Wu}}, \bibinfo {author} {\bibfnamefont {T.}~\bibnamefont {Lovorn}}, \bibinfo
  {author} {\bibfnamefont {E.}~\bibnamefont {Tutuc}}, \bibinfo {author}
  {\bibfnamefont {I.}~\bibnamefont {Martin}},\ and\ \bibinfo {author}
  {\bibfnamefont {A.}~\bibnamefont {MacDonald}},\ }\href@noop {} {\bibfield
  {journal} {\bibinfo  {journal} {Physical review letters}\ }\textbf {\bibinfo
  {volume} {122}},\ \bibinfo {pages} {086402} (\bibinfo {year}
  {2019})}\BibitemShut {NoStop}%
\bibitem [{\citenamefont {Yankowitz}\ \emph {et~al.}(2019)\citenamefont
  {Yankowitz}, \citenamefont {Chen}, \citenamefont {Polshyn}, \citenamefont
  {Zhang}, \citenamefont {Watanabe}, \citenamefont {Taniguchi}, \citenamefont
  {Graf}, \citenamefont {Young},\ and\ \citenamefont
  {Dean}}]{yankowitz2019tuning}%
  \BibitemOpen
  \bibfield  {author} {\bibinfo {author} {\bibfnamefont {M.}~\bibnamefont
  {Yankowitz}}, \bibinfo {author} {\bibfnamefont {S.}~\bibnamefont {Chen}},
  \bibinfo {author} {\bibfnamefont {H.}~\bibnamefont {Polshyn}}, \bibinfo
  {author} {\bibfnamefont {Y.}~\bibnamefont {Zhang}}, \bibinfo {author}
  {\bibfnamefont {K.}~\bibnamefont {Watanabe}}, \bibinfo {author}
  {\bibfnamefont {T.}~\bibnamefont {Taniguchi}}, \bibinfo {author}
  {\bibfnamefont {D.}~\bibnamefont {Graf}}, \bibinfo {author} {\bibfnamefont
  {A.~F.}\ \bibnamefont {Young}},\ and\ \bibinfo {author} {\bibfnamefont
  {C.~R.}\ \bibnamefont {Dean}},\ }\href@noop {} {\bibfield  {journal}
  {\bibinfo  {journal} {Science}\ }\textbf {\bibinfo {volume} {363}},\ \bibinfo
  {pages} {1059} (\bibinfo {year} {2019})}\BibitemShut {NoStop}%
\bibitem [{\citenamefont {Kim}\ \emph {et~al.}(2017)\citenamefont {Kim},
  \citenamefont {DaSilva}, \citenamefont {Huang}, \citenamefont {Fallahazad},
  \citenamefont {Larentis}, \citenamefont {Taniguchi}, \citenamefont
  {Watanabe}, \citenamefont {LeRoy}, \citenamefont {MacDonald},\ and\
  \citenamefont {Tutuc}}]{kim2017tunable}%
  \BibitemOpen
  \bibfield  {author} {\bibinfo {author} {\bibfnamefont {K.}~\bibnamefont
  {Kim}}, \bibinfo {author} {\bibfnamefont {A.}~\bibnamefont {DaSilva}},
  \bibinfo {author} {\bibfnamefont {S.}~\bibnamefont {Huang}}, \bibinfo
  {author} {\bibfnamefont {B.}~\bibnamefont {Fallahazad}}, \bibinfo {author}
  {\bibfnamefont {S.}~\bibnamefont {Larentis}}, \bibinfo {author}
  {\bibfnamefont {T.}~\bibnamefont {Taniguchi}}, \bibinfo {author}
  {\bibfnamefont {K.}~\bibnamefont {Watanabe}}, \bibinfo {author}
  {\bibfnamefont {B.~J.}\ \bibnamefont {LeRoy}}, \bibinfo {author}
  {\bibfnamefont {A.~H.}\ \bibnamefont {MacDonald}},\ and\ \bibinfo {author}
  {\bibfnamefont {E.}~\bibnamefont {Tutuc}},\ }\href@noop {} {\bibfield
  {journal} {\bibinfo  {journal} {Proceedings of the National Academy of
  Sciences}\ }\textbf {\bibinfo {volume} {114}},\ \bibinfo {pages} {3364}
  (\bibinfo {year} {2017})}\BibitemShut {NoStop}%
\bibitem [{\citenamefont {Bistritzer}\ and\ \citenamefont
  {MacDonald}(2011)}]{bistritzer2011moire}%
  \BibitemOpen
  \bibfield  {author} {\bibinfo {author} {\bibfnamefont {R.}~\bibnamefont
  {Bistritzer}}\ and\ \bibinfo {author} {\bibfnamefont {A.~H.}\ \bibnamefont
  {MacDonald}},\ }\href@noop {} {\bibfield  {journal} {\bibinfo  {journal}
  {Proceedings of the National Academy of Sciences}\ }\textbf {\bibinfo
  {volume} {108}},\ \bibinfo {pages} {12233} (\bibinfo {year}
  {2011})}\BibitemShut {NoStop}%
\bibitem [{\citenamefont {Cao}\ \emph {et~al.}(2020)\citenamefont {Cao},
  \citenamefont {Rodan-Legrain}, \citenamefont {Rubies-Bigorda}, \citenamefont
  {Park}, \citenamefont {Watanabe}, \citenamefont {Taniguchi},\ and\
  \citenamefont {Jarillo-Herrero}}]{cao2020tunable}%
  \BibitemOpen
  \bibfield  {author} {\bibinfo {author} {\bibfnamefont {Y.}~\bibnamefont
  {Cao}}, \bibinfo {author} {\bibfnamefont {D.}~\bibnamefont {Rodan-Legrain}},
  \bibinfo {author} {\bibfnamefont {O.}~\bibnamefont {Rubies-Bigorda}},
  \bibinfo {author} {\bibfnamefont {J.~M.}\ \bibnamefont {Park}}, \bibinfo
  {author} {\bibfnamefont {K.}~\bibnamefont {Watanabe}}, \bibinfo {author}
  {\bibfnamefont {T.}~\bibnamefont {Taniguchi}},\ and\ \bibinfo {author}
  {\bibfnamefont {P.}~\bibnamefont {Jarillo-Herrero}},\ }\href@noop {}
  {\bibfield  {journal} {\bibinfo  {journal} {Nature}\ }\textbf {\bibinfo
  {volume} {583}},\ \bibinfo {pages} {215} (\bibinfo {year}
  {2020})}\BibitemShut {NoStop}%
\bibitem [{\citenamefont {Liu}\ \emph {et~al.}(2014)\citenamefont {Liu},
  \citenamefont {Zhang}, \citenamefont {Cao}, \citenamefont {Jin},
  \citenamefont {Qiu}, \citenamefont {Zhou}, \citenamefont {Zettl},
  \citenamefont {Yang}, \citenamefont {Louie},\ and\ \citenamefont
  {Wang}}]{liu2014evolution}%
  \BibitemOpen
  \bibfield  {author} {\bibinfo {author} {\bibfnamefont {K.}~\bibnamefont
  {Liu}}, \bibinfo {author} {\bibfnamefont {L.}~\bibnamefont {Zhang}}, \bibinfo
  {author} {\bibfnamefont {T.}~\bibnamefont {Cao}}, \bibinfo {author}
  {\bibfnamefont {C.}~\bibnamefont {Jin}}, \bibinfo {author} {\bibfnamefont
  {D.}~\bibnamefont {Qiu}}, \bibinfo {author} {\bibfnamefont {Q.}~\bibnamefont
  {Zhou}}, \bibinfo {author} {\bibfnamefont {A.}~\bibnamefont {Zettl}},
  \bibinfo {author} {\bibfnamefont {P.}~\bibnamefont {Yang}}, \bibinfo {author}
  {\bibfnamefont {S.~G.}\ \bibnamefont {Louie}},\ and\ \bibinfo {author}
  {\bibfnamefont {F.}~\bibnamefont {Wang}},\ }\href@noop {} {\bibfield
  {journal} {\bibinfo  {journal} {Nature communications}\ }\textbf {\bibinfo
  {volume} {5}},\ \bibinfo {pages} {4966} (\bibinfo {year} {2014})}\BibitemShut
  {NoStop}%
\bibitem [{\citenamefont {Puretzky}\ \emph {et~al.}(2016)\citenamefont
  {Puretzky}, \citenamefont {Liang}, \citenamefont {Li}, \citenamefont {Xiao},
  \citenamefont {Sumpter}, \citenamefont {Meunier},\ and\ \citenamefont
  {Geohegan}}]{puretzky2016twisted}%
  \BibitemOpen
  \bibfield  {author} {\bibinfo {author} {\bibfnamefont {A.~A.}\ \bibnamefont
  {Puretzky}}, \bibinfo {author} {\bibfnamefont {L.}~\bibnamefont {Liang}},
  \bibinfo {author} {\bibfnamefont {X.}~\bibnamefont {Li}}, \bibinfo {author}
  {\bibfnamefont {K.}~\bibnamefont {Xiao}}, \bibinfo {author} {\bibfnamefont
  {B.~G.}\ \bibnamefont {Sumpter}}, \bibinfo {author} {\bibfnamefont
  {V.}~\bibnamefont {Meunier}},\ and\ \bibinfo {author} {\bibfnamefont {D.~B.}\
  \bibnamefont {Geohegan}},\ }\href@noop {} {\bibfield  {journal} {\bibinfo
  {journal} {ACS nano}\ }\textbf {\bibinfo {volume} {10}},\ \bibinfo {pages}
  {2736} (\bibinfo {year} {2016})}\BibitemShut {NoStop}%
\bibitem [{\citenamefont {Huang}\ \emph {et~al.}(2014)\citenamefont {Huang},
  \citenamefont {Ling}, \citenamefont {Liang}, \citenamefont {Kong},
  \citenamefont {Terrones}, \citenamefont {Meunier},\ and\ \citenamefont
  {Dresselhaus}}]{huang2014probing}%
  \BibitemOpen
  \bibfield  {author} {\bibinfo {author} {\bibfnamefont {S.}~\bibnamefont
  {Huang}}, \bibinfo {author} {\bibfnamefont {X.}~\bibnamefont {Ling}},
  \bibinfo {author} {\bibfnamefont {L.}~\bibnamefont {Liang}}, \bibinfo
  {author} {\bibfnamefont {J.}~\bibnamefont {Kong}}, \bibinfo {author}
  {\bibfnamefont {H.}~\bibnamefont {Terrones}}, \bibinfo {author}
  {\bibfnamefont {V.}~\bibnamefont {Meunier}},\ and\ \bibinfo {author}
  {\bibfnamefont {M.~S.}\ \bibnamefont {Dresselhaus}},\ }\href@noop {}
  {\bibfield  {journal} {\bibinfo  {journal} {Nano letters}\ }\textbf {\bibinfo
  {volume} {14}},\ \bibinfo {pages} {5500} (\bibinfo {year}
  {2014})}\BibitemShut {NoStop}%
\bibitem [{\citenamefont {Wang}\ \emph {et~al.}(2020)\citenamefont {Wang},
  \citenamefont {Shih}, \citenamefont {Ghiotto}, \citenamefont {Xian},
  \citenamefont {Rhodes}, \citenamefont {Tan}, \citenamefont {Claassen},
  \citenamefont {Kennes}, \citenamefont {Bai}, \citenamefont {Kim} \emph
  {et~al.}}]{wang2020correlated}%
  \BibitemOpen
  \bibfield  {author} {\bibinfo {author} {\bibfnamefont {L.}~\bibnamefont
  {Wang}}, \bibinfo {author} {\bibfnamefont {E.-M.}\ \bibnamefont {Shih}},
  \bibinfo {author} {\bibfnamefont {A.}~\bibnamefont {Ghiotto}}, \bibinfo
  {author} {\bibfnamefont {L.}~\bibnamefont {Xian}}, \bibinfo {author}
  {\bibfnamefont {D.~A.}\ \bibnamefont {Rhodes}}, \bibinfo {author}
  {\bibfnamefont {C.}~\bibnamefont {Tan}}, \bibinfo {author} {\bibfnamefont
  {M.}~\bibnamefont {Claassen}}, \bibinfo {author} {\bibfnamefont {D.~M.}\
  \bibnamefont {Kennes}}, \bibinfo {author} {\bibfnamefont {Y.}~\bibnamefont
  {Bai}}, \bibinfo {author} {\bibfnamefont {B.}~\bibnamefont {Kim}}, \emph
  {et~al.},\ }\href@noop {} {\bibfield  {journal} {\bibinfo  {journal} {Nature
  materials}\ }\textbf {\bibinfo {volume} {19}},\ \bibinfo {pages} {861}
  (\bibinfo {year} {2020})}\BibitemShut {NoStop}%
\bibitem [{\citenamefont {Morell}\ \emph {et~al.}(2010)\citenamefont {Morell},
  \citenamefont {Correa}, \citenamefont {Vargas}, \citenamefont {Pacheco},\
  and\ \citenamefont {Barticevic}}]{morell2010flat}%
  \BibitemOpen
  \bibfield  {author} {\bibinfo {author} {\bibfnamefont {E.~S.}\ \bibnamefont
  {Morell}}, \bibinfo {author} {\bibfnamefont {J.}~\bibnamefont {Correa}},
  \bibinfo {author} {\bibfnamefont {P.}~\bibnamefont {Vargas}}, \bibinfo
  {author} {\bibfnamefont {M.}~\bibnamefont {Pacheco}},\ and\ \bibinfo {author}
  {\bibfnamefont {Z.}~\bibnamefont {Barticevic}},\ }\href@noop {} {\bibfield
  {journal} {\bibinfo  {journal} {Physical Review B}\ }\textbf {\bibinfo
  {volume} {82}},\ \bibinfo {pages} {121407} (\bibinfo {year}
  {2010})}\BibitemShut {NoStop}%
\bibitem [{\citenamefont {Zhang}\ \emph {et~al.}(2020)\citenamefont {Zhang},
  \citenamefont {Wang}, \citenamefont {Watanabe}, \citenamefont {Taniguchi},
  \citenamefont {Ueno}, \citenamefont {Tutuc},\ and\ \citenamefont
  {LeRoy}}]{zhang2020flat}%
  \BibitemOpen
  \bibfield  {author} {\bibinfo {author} {\bibfnamefont {Z.}~\bibnamefont
  {Zhang}}, \bibinfo {author} {\bibfnamefont {Y.}~\bibnamefont {Wang}},
  \bibinfo {author} {\bibfnamefont {K.}~\bibnamefont {Watanabe}}, \bibinfo
  {author} {\bibfnamefont {T.}~\bibnamefont {Taniguchi}}, \bibinfo {author}
  {\bibfnamefont {K.}~\bibnamefont {Ueno}}, \bibinfo {author} {\bibfnamefont
  {E.}~\bibnamefont {Tutuc}},\ and\ \bibinfo {author} {\bibfnamefont {B.~J.}\
  \bibnamefont {LeRoy}},\ }\href@noop {} {\bibfield  {journal} {\bibinfo
  {journal} {Nature Physics}\ }\textbf {\bibinfo {volume} {16}},\ \bibinfo
  {pages} {1093} (\bibinfo {year} {2020})}\BibitemShut {NoStop}%
\bibitem [{\citenamefont {Regan}\ \emph {et~al.}(2020)\citenamefont {Regan},
  \citenamefont {Wang}, \citenamefont {Jin}, \citenamefont {Bakti~Utama},
  \citenamefont {Gao}, \citenamefont {Wei}, \citenamefont {Zhao}, \citenamefont
  {Zhao}, \citenamefont {Zhang}, \citenamefont {Yumigeta}, \citenamefont
  {Blei}, \citenamefont {Carlstr{\"o}m}, \citenamefont {Watanabe},
  \citenamefont {Taniguchi}, \citenamefont {Tongay}, \citenamefont {Crommie},
  \citenamefont {Zettl},\ and\ \citenamefont {Wang}}]{Regan2020-po}%
  \BibitemOpen
  \bibfield  {author} {\bibinfo {author} {\bibfnamefont {E.~C.}\ \bibnamefont
  {Regan}}, \bibinfo {author} {\bibfnamefont {D.}~\bibnamefont {Wang}},
  \bibinfo {author} {\bibfnamefont {C.}~\bibnamefont {Jin}}, \bibinfo {author}
  {\bibfnamefont {M.~I.}\ \bibnamefont {Bakti~Utama}}, \bibinfo {author}
  {\bibfnamefont {B.}~\bibnamefont {Gao}}, \bibinfo {author} {\bibfnamefont
  {X.}~\bibnamefont {Wei}}, \bibinfo {author} {\bibfnamefont {S.}~\bibnamefont
  {Zhao}}, \bibinfo {author} {\bibfnamefont {W.}~\bibnamefont {Zhao}}, \bibinfo
  {author} {\bibfnamefont {Z.}~\bibnamefont {Zhang}}, \bibinfo {author}
  {\bibfnamefont {K.}~\bibnamefont {Yumigeta}}, \bibinfo {author}
  {\bibfnamefont {M.}~\bibnamefont {Blei}}, \bibinfo {author} {\bibfnamefont
  {J.~D.}\ \bibnamefont {Carlstr{\"o}m}}, \bibinfo {author} {\bibfnamefont
  {K.}~\bibnamefont {Watanabe}}, \bibinfo {author} {\bibfnamefont
  {T.}~\bibnamefont {Taniguchi}}, \bibinfo {author} {\bibfnamefont
  {S.}~\bibnamefont {Tongay}}, \bibinfo {author} {\bibfnamefont
  {M.}~\bibnamefont {Crommie}}, \bibinfo {author} {\bibfnamefont
  {A.}~\bibnamefont {Zettl}},\ and\ \bibinfo {author} {\bibfnamefont
  {F.}~\bibnamefont {Wang}},\ }\href@noop {} {\bibfield  {journal} {\bibinfo
  {journal} {Nature}\ }\textbf {\bibinfo {volume} {579}},\ \bibinfo {pages}
  {359} (\bibinfo {year} {2020})}\BibitemShut {NoStop}%
\bibitem [{\citenamefont {Tang}\ \emph {et~al.}(2020)\citenamefont {Tang},
  \citenamefont {Li}, \citenamefont {Li}, \citenamefont {Xu}, \citenamefont
  {Liu}, \citenamefont {Barmak}, \citenamefont {Watanabe}, \citenamefont
  {Taniguchi}, \citenamefont {MacDonald}, \citenamefont {Shan},\ and\
  \citenamefont {Mak}}]{Tang2020-pm}%
  \BibitemOpen
  \bibfield  {author} {\bibinfo {author} {\bibfnamefont {Y.}~\bibnamefont
  {Tang}}, \bibinfo {author} {\bibfnamefont {L.}~\bibnamefont {Li}}, \bibinfo
  {author} {\bibfnamefont {T.}~\bibnamefont {Li}}, \bibinfo {author}
  {\bibfnamefont {Y.}~\bibnamefont {Xu}}, \bibinfo {author} {\bibfnamefont
  {S.}~\bibnamefont {Liu}}, \bibinfo {author} {\bibfnamefont {K.}~\bibnamefont
  {Barmak}}, \bibinfo {author} {\bibfnamefont {K.}~\bibnamefont {Watanabe}},
  \bibinfo {author} {\bibfnamefont {T.}~\bibnamefont {Taniguchi}}, \bibinfo
  {author} {\bibfnamefont {A.~H.}\ \bibnamefont {MacDonald}}, \bibinfo {author}
  {\bibfnamefont {J.}~\bibnamefont {Shan}},\ and\ \bibinfo {author}
  {\bibfnamefont {K.~F.}\ \bibnamefont {Mak}},\ }\href@noop {} {\bibfield
  {journal} {\bibinfo  {journal} {Nature}\ }\textbf {\bibinfo {volume} {579}},\
  \bibinfo {pages} {353} (\bibinfo {year} {2020})}\BibitemShut {NoStop}%
\bibitem [{\citenamefont {Park}\ \emph
  {et~al.}(2023{\natexlab{a}})\citenamefont {Park}, \citenamefont {Cai},
  \citenamefont {Anderson}, \citenamefont {Zhang}, \citenamefont {Zhu},
  \citenamefont {Liu}, \citenamefont {Wang}, \citenamefont {Holtzmann},
  \citenamefont {Hu}, \citenamefont {Liu}, \citenamefont {Taniguchi},
  \citenamefont {Watanabe}, \citenamefont {Chu}, \citenamefont {Cao},
  \citenamefont {Fu}, \citenamefont {Yao}, \citenamefont {Chang}, \citenamefont
  {Cobden}, \citenamefont {Xiao},\ and\ \citenamefont {Xu}}]{Park2023-an}%
  \BibitemOpen
  \bibfield  {author} {\bibinfo {author} {\bibfnamefont {H.}~\bibnamefont
  {Park}}, \bibinfo {author} {\bibfnamefont {J.}~\bibnamefont {Cai}}, \bibinfo
  {author} {\bibfnamefont {E.}~\bibnamefont {Anderson}}, \bibinfo {author}
  {\bibfnamefont {Y.}~\bibnamefont {Zhang}}, \bibinfo {author} {\bibfnamefont
  {J.}~\bibnamefont {Zhu}}, \bibinfo {author} {\bibfnamefont {X.}~\bibnamefont
  {Liu}}, \bibinfo {author} {\bibfnamefont {C.}~\bibnamefont {Wang}}, \bibinfo
  {author} {\bibfnamefont {W.}~\bibnamefont {Holtzmann}}, \bibinfo {author}
  {\bibfnamefont {C.}~\bibnamefont {Hu}}, \bibinfo {author} {\bibfnamefont
  {Z.}~\bibnamefont {Liu}}, \bibinfo {author} {\bibfnamefont {T.}~\bibnamefont
  {Taniguchi}}, \bibinfo {author} {\bibfnamefont {K.}~\bibnamefont {Watanabe}},
  \bibinfo {author} {\bibfnamefont {J.-H.}\ \bibnamefont {Chu}}, \bibinfo
  {author} {\bibfnamefont {T.}~\bibnamefont {Cao}}, \bibinfo {author}
  {\bibfnamefont {L.}~\bibnamefont {Fu}}, \bibinfo {author} {\bibfnamefont
  {W.}~\bibnamefont {Yao}}, \bibinfo {author} {\bibfnamefont {C.-Z.}\
  \bibnamefont {Chang}}, \bibinfo {author} {\bibfnamefont {D.}~\bibnamefont
  {Cobden}}, \bibinfo {author} {\bibfnamefont {D.}~\bibnamefont {Xiao}},\ and\
  \bibinfo {author} {\bibfnamefont {X.}~\bibnamefont {Xu}},\ }\href@noop {}
  {\bibfield  {journal} {\bibinfo  {journal} {Nature}\ }\textbf {\bibinfo
  {volume} {622}},\ \bibinfo {pages} {74} (\bibinfo {year}
  {2023}{\natexlab{a}})}\BibitemShut {NoStop}%
\bibitem [{\citenamefont {Kang}\ \emph
  {et~al.}(2024{\natexlab{a}})\citenamefont {Kang}, \citenamefont {Shen},
  \citenamefont {Qiu}, \citenamefont {Zeng}, \citenamefont {Xia}, \citenamefont
  {Watanabe}, \citenamefont {Taniguchi}, \citenamefont {Shan},\ and\
  \citenamefont {Mak}}]{Kang2024-ig}%
  \BibitemOpen
  \bibfield  {author} {\bibinfo {author} {\bibfnamefont {K.}~\bibnamefont
  {Kang}}, \bibinfo {author} {\bibfnamefont {B.}~\bibnamefont {Shen}}, \bibinfo
  {author} {\bibfnamefont {Y.}~\bibnamefont {Qiu}}, \bibinfo {author}
  {\bibfnamefont {Y.}~\bibnamefont {Zeng}}, \bibinfo {author} {\bibfnamefont
  {Z.}~\bibnamefont {Xia}}, \bibinfo {author} {\bibfnamefont {K.}~\bibnamefont
  {Watanabe}}, \bibinfo {author} {\bibfnamefont {T.}~\bibnamefont {Taniguchi}},
  \bibinfo {author} {\bibfnamefont {J.}~\bibnamefont {Shan}},\ and\ \bibinfo
  {author} {\bibfnamefont {K.~F.}\ \bibnamefont {Mak}},\ }\href@noop {}
  {\bibfield  {journal} {\bibinfo  {journal} {Nature}\ }\textbf {\bibinfo
  {volume} {628}},\ \bibinfo {pages} {522} (\bibinfo {year}
  {2024}{\natexlab{a}})}\BibitemShut {NoStop}%
\bibitem [{\citenamefont {Ji}\ \emph {et~al.}(2024)\citenamefont {Ji},
  \citenamefont {Park}, \citenamefont {Barber}, \citenamefont {Hu},
  \citenamefont {Watanabe}, \citenamefont {Taniguchi}, \citenamefont {Chu},
  \citenamefont {Xu},\ and\ \citenamefont {Shen}}]{Ji2024-uy}%
  \BibitemOpen
  \bibfield  {author} {\bibinfo {author} {\bibfnamefont {Z.}~\bibnamefont
  {Ji}}, \bibinfo {author} {\bibfnamefont {H.}~\bibnamefont {Park}}, \bibinfo
  {author} {\bibfnamefont {M.~E.}\ \bibnamefont {Barber}}, \bibinfo {author}
  {\bibfnamefont {C.}~\bibnamefont {Hu}}, \bibinfo {author} {\bibfnamefont
  {K.}~\bibnamefont {Watanabe}}, \bibinfo {author} {\bibfnamefont
  {T.}~\bibnamefont {Taniguchi}}, \bibinfo {author} {\bibfnamefont {J.-H.}\
  \bibnamefont {Chu}}, \bibinfo {author} {\bibfnamefont {X.}~\bibnamefont
  {Xu}},\ and\ \bibinfo {author} {\bibfnamefont {Z.-X.}\ \bibnamefont {Shen}},\
  }\href@noop {} {\bibfield  {journal} {\bibinfo  {journal} {Nature}\ }\textbf
  {\bibinfo {volume} {635}},\ \bibinfo {pages} {578} (\bibinfo {year}
  {2024})}\BibitemShut {NoStop}%
\bibitem [{\citenamefont {Kang}\ \emph
  {et~al.}(2024{\natexlab{b}})\citenamefont {Kang}, \citenamefont {Shen},
  \citenamefont {Qiu}, \citenamefont {Zeng}, \citenamefont {Xia}, \citenamefont
  {Watanabe}, \citenamefont {Taniguchi}, \citenamefont {Shan},\ and\
  \citenamefont {Mak}}]{Kang2024-wj}%
  \BibitemOpen
  \bibfield  {author} {\bibinfo {author} {\bibfnamefont {K.}~\bibnamefont
  {Kang}}, \bibinfo {author} {\bibfnamefont {B.}~\bibnamefont {Shen}}, \bibinfo
  {author} {\bibfnamefont {Y.}~\bibnamefont {Qiu}}, \bibinfo {author}
  {\bibfnamefont {Y.}~\bibnamefont {Zeng}}, \bibinfo {author} {\bibfnamefont
  {Z.}~\bibnamefont {Xia}}, \bibinfo {author} {\bibfnamefont {K.}~\bibnamefont
  {Watanabe}}, \bibinfo {author} {\bibfnamefont {T.}~\bibnamefont {Taniguchi}},
  \bibinfo {author} {\bibfnamefont {J.}~\bibnamefont {Shan}},\ and\ \bibinfo
  {author} {\bibfnamefont {K.~F.}\ \bibnamefont {Mak}},\ }\href@noop {}
  {\bibfield  {journal} {\bibinfo  {journal} {Nature}\ }\textbf {\bibinfo
  {volume} {628}},\ \bibinfo {pages} {522} (\bibinfo {year}
  {2024}{\natexlab{b}})}\BibitemShut {NoStop}%
\bibitem [{\citenamefont {Park}\ \emph
  {et~al.}(2023{\natexlab{b}})\citenamefont {Park}, \citenamefont {Cai},
  \citenamefont {Anderson}, \citenamefont {Zhang}, \citenamefont {Zhu},
  \citenamefont {Liu}, \citenamefont {Wang}, \citenamefont {Holtzmann},
  \citenamefont {Hu}, \citenamefont {Liu}, \citenamefont {Taniguchi},
  \citenamefont {Watanabe}, \citenamefont {Chu}, \citenamefont {Cao},
  \citenamefont {Fu}, \citenamefont {Yao}, \citenamefont {Chang}, \citenamefont
  {Cobden}, \citenamefont {Xiao},\ and\ \citenamefont {Xu}}]{Park2023-xg}%
  \BibitemOpen
  \bibfield  {author} {\bibinfo {author} {\bibfnamefont {H.}~\bibnamefont
  {Park}}, \bibinfo {author} {\bibfnamefont {J.}~\bibnamefont {Cai}}, \bibinfo
  {author} {\bibfnamefont {E.}~\bibnamefont {Anderson}}, \bibinfo {author}
  {\bibfnamefont {Y.}~\bibnamefont {Zhang}}, \bibinfo {author} {\bibfnamefont
  {J.}~\bibnamefont {Zhu}}, \bibinfo {author} {\bibfnamefont {X.}~\bibnamefont
  {Liu}}, \bibinfo {author} {\bibfnamefont {C.}~\bibnamefont {Wang}}, \bibinfo
  {author} {\bibfnamefont {W.}~\bibnamefont {Holtzmann}}, \bibinfo {author}
  {\bibfnamefont {C.}~\bibnamefont {Hu}}, \bibinfo {author} {\bibfnamefont
  {Z.}~\bibnamefont {Liu}}, \bibinfo {author} {\bibfnamefont {T.}~\bibnamefont
  {Taniguchi}}, \bibinfo {author} {\bibfnamefont {K.}~\bibnamefont {Watanabe}},
  \bibinfo {author} {\bibfnamefont {J.-H.}\ \bibnamefont {Chu}}, \bibinfo
  {author} {\bibfnamefont {T.}~\bibnamefont {Cao}}, \bibinfo {author}
  {\bibfnamefont {L.}~\bibnamefont {Fu}}, \bibinfo {author} {\bibfnamefont
  {W.}~\bibnamefont {Yao}}, \bibinfo {author} {\bibfnamefont {C.-Z.}\
  \bibnamefont {Chang}}, \bibinfo {author} {\bibfnamefont {D.}~\bibnamefont
  {Cobden}}, \bibinfo {author} {\bibfnamefont {D.}~\bibnamefont {Xiao}},\ and\
  \bibinfo {author} {\bibfnamefont {X.}~\bibnamefont {Xu}},\ }\href@noop {}
  {\bibfield  {journal} {\bibinfo  {journal} {Nature}\ }\textbf {\bibinfo
  {volume} {622}},\ \bibinfo {pages} {74} (\bibinfo {year}
  {2023}{\natexlab{b}})}\BibitemShut {NoStop}%
\bibitem [{\citenamefont {Holmes}\ \emph {et~al.}(2015)\citenamefont {Holmes},
  \citenamefont {Iuliucci}, \citenamefont {Mueller},\ and\ \citenamefont
  {Dybowski}}]{Holmes2015-jb}%
  \BibitemOpen
  \bibfield  {author} {\bibinfo {author} {\bibfnamefont {S.~T.}\ \bibnamefont
  {Holmes}}, \bibinfo {author} {\bibfnamefont {R.~J.}\ \bibnamefont
  {Iuliucci}}, \bibinfo {author} {\bibfnamefont {K.~T.}\ \bibnamefont
  {Mueller}},\ and\ \bibinfo {author} {\bibfnamefont {C.}~\bibnamefont
  {Dybowski}},\ }\href@noop {} {\bibfield  {journal} {\bibinfo  {journal} {J.
  Chem. Theory Comput.}\ }\textbf {\bibinfo {volume} {11}},\ \bibinfo {pages}
  {5229} (\bibinfo {year} {2015})}\BibitemShut {NoStop}%
\bibitem [{\citenamefont {Iv{\'a}dy}\ \emph {et~al.}(2013)\citenamefont
  {Iv{\'a}dy}, \citenamefont {Abrikosov}, \citenamefont {Janz{\'e}n},\ and\
  \citenamefont {Gali}}]{ivady2013role}%
  \BibitemOpen
  \bibfield  {author} {\bibinfo {author} {\bibfnamefont {V.}~\bibnamefont
  {Iv{\'a}dy}}, \bibinfo {author} {\bibfnamefont {I.}~\bibnamefont
  {Abrikosov}}, \bibinfo {author} {\bibfnamefont {E.}~\bibnamefont
  {Janz{\'e}n}},\ and\ \bibinfo {author} {\bibfnamefont {A.}~\bibnamefont
  {Gali}},\ }\href@noop {} {\bibfield  {journal} {\bibinfo  {journal} {Physical
  Review B—Condensed Matter and Materials Physics}\ }\textbf {\bibinfo
  {volume} {87}},\ \bibinfo {pages} {205201} (\bibinfo {year}
  {2013})}\BibitemShut {NoStop}%
\bibitem [{\citenamefont {Oh}\ \emph {et~al.}(2024)\citenamefont {Oh},
  \citenamefont {Song},\ and\ \citenamefont {Bae}}]{oh2024review}%
  \BibitemOpen
  \bibfield  {author} {\bibinfo {author} {\bibfnamefont {Y.}~\bibnamefont
  {Oh}}, \bibinfo {author} {\bibfnamefont {S.}~\bibnamefont {Song}},\ and\
  \bibinfo {author} {\bibfnamefont {J.}~\bibnamefont {Bae}},\ }\href@noop {}
  {\bibfield  {journal} {\bibinfo  {journal} {International Journal of
  Molecular Sciences}\ }\textbf {\bibinfo {volume} {25}},\ \bibinfo {pages}
  {13104} (\bibinfo {year} {2024})}\BibitemShut {NoStop}%
\bibitem [{\citenamefont {Vinh}\ \emph {et~al.}(2025)\citenamefont {Vinh},
  \citenamefont {Lu},\ and\ \citenamefont {Pham}}]{vinh2025stacking}%
  \BibitemOpen
  \bibfield  {author} {\bibinfo {author} {\bibfnamefont {N.~V.}\ \bibnamefont
  {Vinh}}, \bibinfo {author} {\bibfnamefont {D.}~\bibnamefont {Lu}},\ and\
  \bibinfo {author} {\bibfnamefont {K.}~\bibnamefont {Pham}},\ }\href@noop {}
  {\bibfield  {journal} {\bibinfo  {journal} {Nanoscale Advances}\ } (\bibinfo
  {year} {2025})}\BibitemShut {NoStop}%
\bibitem [{\citenamefont {Zhou}\ \emph {et~al.}(2015)\citenamefont {Zhou},
  \citenamefont {Han}, \citenamefont {Dai}, \citenamefont {Sun},\ and\
  \citenamefont {Srolovitz}}]{zhou2015van}%
  \BibitemOpen
  \bibfield  {author} {\bibinfo {author} {\bibfnamefont {S.}~\bibnamefont
  {Zhou}}, \bibinfo {author} {\bibfnamefont {J.}~\bibnamefont {Han}}, \bibinfo
  {author} {\bibfnamefont {S.}~\bibnamefont {Dai}}, \bibinfo {author}
  {\bibfnamefont {J.}~\bibnamefont {Sun}},\ and\ \bibinfo {author}
  {\bibfnamefont {D.~J.}\ \bibnamefont {Srolovitz}},\ }\href@noop {} {\bibfield
   {journal} {\bibinfo  {journal} {Physical Review B}\ }\textbf {\bibinfo
  {volume} {92}},\ \bibinfo {pages} {155438} (\bibinfo {year}
  {2015})}\BibitemShut {NoStop}%
\bibitem [{\citenamefont {Grundmann}(2010)}]{grundmann2010physics}%
  \BibitemOpen
  \bibfield  {author} {\bibinfo {author} {\bibfnamefont {M.}~\bibnamefont
  {Grundmann}},\ }\href@noop {} {\emph {\bibinfo {title} {Physics of
  semiconductors}}},\ Vol.~\bibinfo {volume} {11}\ (\bibinfo  {publisher}
  {Springer},\ \bibinfo {year} {2010})\ pp.\ \bibinfo {pages}
  {401--472}\BibitemShut {NoStop}%
\bibitem [{\citenamefont {Sze}\ \emph {et~al.}(2021)\citenamefont {Sze},
  \citenamefont {Li},\ and\ \citenamefont {Ng}}]{sze2021physics}%
  \BibitemOpen
  \bibfield  {author} {\bibinfo {author} {\bibfnamefont {S.~M.}\ \bibnamefont
  {Sze}}, \bibinfo {author} {\bibfnamefont {Y.}~\bibnamefont {Li}},\ and\
  \bibinfo {author} {\bibfnamefont {K.~K.}\ \bibnamefont {Ng}},\ }\href@noop {}
  {\emph {\bibinfo {title} {Physics of semiconductor devices}}}\ (\bibinfo
  {publisher} {John wiley \& sons},\ \bibinfo {year} {2021})\ pp.\ \bibinfo
  {pages} {697--714}\BibitemShut {NoStop}%
\bibitem [{\citenamefont {Singh}(2000)}]{singh2000semiconductor}%
  \BibitemOpen
  \bibfield  {author} {\bibinfo {author} {\bibfnamefont {J.}~\bibnamefont
  {Singh}},\ }\href@noop {} {\emph {\bibinfo {title} {Semiconductor devices:
  basic principles}}}\ (\bibinfo  {publisher} {John Wiley \& Sons},\ \bibinfo
  {year} {2000})\ pp.\ \bibinfo {pages} {460--502}\BibitemShut {NoStop}%
\bibitem [{\citenamefont {Yuan}\ \emph {et~al.}(2018)\citenamefont {Yuan},
  \citenamefont {Deng}, \citenamefont {Li}, \citenamefont {Wei},\ and\
  \citenamefont {Luo}}]{yuan2018unified}%
  \BibitemOpen
  \bibfield  {author} {\bibinfo {author} {\bibfnamefont {L.-D.}\ \bibnamefont
  {Yuan}}, \bibinfo {author} {\bibfnamefont {H.-X.}\ \bibnamefont {Deng}},
  \bibinfo {author} {\bibfnamefont {S.-S.}\ \bibnamefont {Li}}, \bibinfo
  {author} {\bibfnamefont {S.-H.}\ \bibnamefont {Wei}},\ and\ \bibinfo {author}
  {\bibfnamefont {J.-W.}\ \bibnamefont {Luo}},\ }\href@noop {} {\bibfield
  {journal} {\bibinfo  {journal} {Physical Review B}\ }\textbf {\bibinfo
  {volume} {98}},\ \bibinfo {pages} {245203} (\bibinfo {year}
  {2018})}\BibitemShut {NoStop}%
\bibitem [{\citenamefont {Necio}\ and\ \citenamefont
  {Birowska}(2020)}]{necio2020supercell}%
  \BibitemOpen
  \bibfield  {author} {\bibinfo {author} {\bibfnamefont {T.}~\bibnamefont
  {Necio}}\ and\ \bibinfo {author} {\bibfnamefont {M.}~\bibnamefont
  {Birowska}},\ }\href@noop {} {\bibfield  {journal} {\bibinfo  {journal} {AIP
  Advances}\ }\textbf {\bibinfo {volume} {10}},\ \bibinfo {pages} {105105}
  (\bibinfo {year} {2020})}\BibitemShut {NoStop}%
\bibitem [{\citenamefont {Zur}\ and\ \citenamefont {McGill}(1984)}]{Zur1984}%
  \BibitemOpen
  \bibfield  {author} {\bibinfo {author} {\bibfnamefont {A.}~\bibnamefont
  {Zur}}\ and\ \bibinfo {author} {\bibfnamefont {T.~C.}\ \bibnamefont
  {McGill}},\ }\href {https://doi.org/10.1063/1.333084} {\bibfield  {journal}
  {\bibinfo  {journal} {Journal of Applied Physics}\ }\textbf {\bibinfo
  {volume} {55}},\ \bibinfo {pages} {378} (\bibinfo {year} {1984})}\BibitemShut
  {NoStop}%
\bibitem [{\citenamefont {Stokes}\ and\ \citenamefont
  {Hatch}(2005)}]{stokes2005findsym}%
  \BibitemOpen
  \bibfield  {author} {\bibinfo {author} {\bibfnamefont {H.~T.}\ \bibnamefont
  {Stokes}}\ and\ \bibinfo {author} {\bibfnamefont {D.~M.}\ \bibnamefont
  {Hatch}},\ }\href@noop {} {\bibfield  {journal} {\bibinfo  {journal} {Journal
  of Applied Crystallography}\ }\textbf {\bibinfo {volume} {38}},\ \bibinfo
  {pages} {237} (\bibinfo {year} {2005})}\BibitemShut {NoStop}%
\bibitem [{\citenamefont {Carr}\ \emph {et~al.}(2017)\citenamefont {Carr},
  \citenamefont {Massatt}, \citenamefont {Fang}, \citenamefont {Cazeaux},
  \citenamefont {Luskin},\ and\ \citenamefont {Kaxiras}}]{carr2017twistronics}%
  \BibitemOpen
  \bibfield  {author} {\bibinfo {author} {\bibfnamefont {S.}~\bibnamefont
  {Carr}}, \bibinfo {author} {\bibfnamefont {D.}~\bibnamefont {Massatt}},
  \bibinfo {author} {\bibfnamefont {S.}~\bibnamefont {Fang}}, \bibinfo {author}
  {\bibfnamefont {P.}~\bibnamefont {Cazeaux}}, \bibinfo {author} {\bibfnamefont
  {M.}~\bibnamefont {Luskin}},\ and\ \bibinfo {author} {\bibfnamefont
  {E.}~\bibnamefont {Kaxiras}},\ }\href@noop {} {\bibfield  {journal} {\bibinfo
   {journal} {Physical Review B}\ }\textbf {\bibinfo {volume} {95}},\ \bibinfo
  {pages} {075420} (\bibinfo {year} {2017})}\BibitemShut {NoStop}%
\bibitem [{\citenamefont {Leconte}\ \emph {et~al.}(2022)\citenamefont
  {Leconte}, \citenamefont {Javvaji}, \citenamefont {An}, \citenamefont
  {Samudrala},\ and\ \citenamefont {Jung}}]{leconte2022relaxation}%
  \BibitemOpen
  \bibfield  {author} {\bibinfo {author} {\bibfnamefont {N.}~\bibnamefont
  {Leconte}}, \bibinfo {author} {\bibfnamefont {S.}~\bibnamefont {Javvaji}},
  \bibinfo {author} {\bibfnamefont {J.}~\bibnamefont {An}}, \bibinfo {author}
  {\bibfnamefont {A.}~\bibnamefont {Samudrala}},\ and\ \bibinfo {author}
  {\bibfnamefont {J.}~\bibnamefont {Jung}},\ }\href@noop {} {\bibfield
  {journal} {\bibinfo  {journal} {Physical Review B}\ }\textbf {\bibinfo
  {volume} {106}},\ \bibinfo {pages} {115410} (\bibinfo {year}
  {2022})}\BibitemShut {NoStop}%
\bibitem [{\citenamefont {Heyd}\ \emph {et~al.}(2003)\citenamefont {Heyd},
  \citenamefont {Scuseria},\ and\ \citenamefont {Ernzerhof}}]{heyd2003hybrid}%
  \BibitemOpen
  \bibfield  {author} {\bibinfo {author} {\bibfnamefont {J.}~\bibnamefont
  {Heyd}}, \bibinfo {author} {\bibfnamefont {G.~E.}\ \bibnamefont {Scuseria}},\
  and\ \bibinfo {author} {\bibfnamefont {M.}~\bibnamefont {Ernzerhof}},\
  }\href@noop {} {\bibfield  {journal} {\bibinfo  {journal} {The Journal of
  chemical physics}\ }\textbf {\bibinfo {volume} {118}},\ \bibinfo {pages}
  {8207} (\bibinfo {year} {2003})}\BibitemShut {NoStop}%
\bibitem [{\citenamefont {Krukau}\ \emph {et~al.}(2006)\citenamefont {Krukau},
  \citenamefont {Vydrov}, \citenamefont {Izmaylov},\ and\ \citenamefont
  {Scuseria}}]{krukau2006influence}%
  \BibitemOpen
  \bibfield  {author} {\bibinfo {author} {\bibfnamefont {A.~V.}\ \bibnamefont
  {Krukau}}, \bibinfo {author} {\bibfnamefont {O.~A.}\ \bibnamefont {Vydrov}},
  \bibinfo {author} {\bibfnamefont {A.~F.}\ \bibnamefont {Izmaylov}},\ and\
  \bibinfo {author} {\bibfnamefont {G.~E.}\ \bibnamefont {Scuseria}},\
  }\href@noop {} {\bibfield  {journal} {\bibinfo  {journal} {The Journal of
  chemical physics}\ }\textbf {\bibinfo {volume} {125}},\ \bibinfo {pages}
  {224106} (\bibinfo {year} {2006})}\BibitemShut {NoStop}%
\bibitem [{\citenamefont {Dovesi}\ \emph {et~al.}(2018)\citenamefont {Dovesi},
  \citenamefont {Erba}, \citenamefont {Orlando}, \citenamefont
  {Zicovich-Wilson}, \citenamefont {Civalleri}, \citenamefont {Maschio},
  \citenamefont {R{\'e}rat}, \citenamefont {Casassa}, \citenamefont {Baima},
  \citenamefont {Salustro} \emph {et~al.}}]{dovesi2018quantum}%
  \BibitemOpen
  \bibfield  {author} {\bibinfo {author} {\bibfnamefont {R.}~\bibnamefont
  {Dovesi}}, \bibinfo {author} {\bibfnamefont {A.}~\bibnamefont {Erba}},
  \bibinfo {author} {\bibfnamefont {R.}~\bibnamefont {Orlando}}, \bibinfo
  {author} {\bibfnamefont {C.~M.}\ \bibnamefont {Zicovich-Wilson}}, \bibinfo
  {author} {\bibfnamefont {B.}~\bibnamefont {Civalleri}}, \bibinfo {author}
  {\bibfnamefont {L.}~\bibnamefont {Maschio}}, \bibinfo {author} {\bibfnamefont
  {M.}~\bibnamefont {R{\'e}rat}}, \bibinfo {author} {\bibfnamefont
  {S.}~\bibnamefont {Casassa}}, \bibinfo {author} {\bibfnamefont
  {J.}~\bibnamefont {Baima}}, \bibinfo {author} {\bibfnamefont
  {S.}~\bibnamefont {Salustro}}, \emph {et~al.},\ }\href@noop {} {\bibfield
  {journal} {\bibinfo  {journal} {Wiley Interdisciplinary Reviews:
  Computational Molecular Science}\ }\textbf {\bibinfo {volume} {8}},\ \bibinfo
  {pages} {e1360} (\bibinfo {year} {2018})}\BibitemShut {NoStop}%
\bibitem [{\citenamefont {Grimme}\ \emph {et~al.}(2010)\citenamefont {Grimme},
  \citenamefont {Antony}, \citenamefont {Ehrlich},\ and\ \citenamefont
  {Krieg}}]{grimme2010consistent}%
  \BibitemOpen
  \bibfield  {author} {\bibinfo {author} {\bibfnamefont {S.}~\bibnamefont
  {Grimme}}, \bibinfo {author} {\bibfnamefont {J.}~\bibnamefont {Antony}},
  \bibinfo {author} {\bibfnamefont {S.}~\bibnamefont {Ehrlich}},\ and\ \bibinfo
  {author} {\bibfnamefont {H.}~\bibnamefont {Krieg}},\ }\href@noop {}
  {\bibfield  {journal} {\bibinfo  {journal} {The Journal of chemical physics}\
  }\textbf {\bibinfo {volume} {132}},\ \bibinfo {pages} {154104} (\bibinfo
  {year} {2010})}\BibitemShut {NoStop}%
\bibitem [{\citenamefont {Becke}\ and\ \citenamefont
  {Johnson}(2005)}]{becke2005density}%
  \BibitemOpen
  \bibfield  {author} {\bibinfo {author} {\bibfnamefont {A.~D.}\ \bibnamefont
  {Becke}}\ and\ \bibinfo {author} {\bibfnamefont {E.~R.}\ \bibnamefont
  {Johnson}},\ }\href@noop {} {\bibfield  {journal} {\bibinfo  {journal} {The
  Journal of chemical physics}\ }\textbf {\bibinfo {volume} {123}},\ \bibinfo
  {pages} {154101} (\bibinfo {year} {2005})}\BibitemShut {NoStop}%
\bibitem [{\citenamefont {Vilela~Oliveira}\ \emph {et~al.}(2019)\citenamefont
  {Vilela~Oliveira}, \citenamefont {Laun}, \citenamefont {Peintinger},\ and\
  \citenamefont {Bredow}}]{vilela2019bsse}%
  \BibitemOpen
  \bibfield  {author} {\bibinfo {author} {\bibfnamefont {D.}~\bibnamefont
  {Vilela~Oliveira}}, \bibinfo {author} {\bibfnamefont {J.}~\bibnamefont
  {Laun}}, \bibinfo {author} {\bibfnamefont {M.~F.}\ \bibnamefont
  {Peintinger}},\ and\ \bibinfo {author} {\bibfnamefont {T.}~\bibnamefont
  {Bredow}},\ }\href@noop {} {\bibfield  {journal} {\bibinfo  {journal}
  {Journal of Computational Chemistry}\ }\textbf {\bibinfo {volume} {40}},\
  \bibinfo {pages} {2364} (\bibinfo {year} {2019})}\BibitemShut {NoStop}%
\bibitem [{\citenamefont {Monkhorst}\ and\ \citenamefont
  {Pack}(1976)}]{monkhorst1976special}%
  \BibitemOpen
  \bibfield  {author} {\bibinfo {author} {\bibfnamefont {H.~J.}\ \bibnamefont
  {Monkhorst}}\ and\ \bibinfo {author} {\bibfnamefont {J.~D.}\ \bibnamefont
  {Pack}},\ }\href@noop {} {\bibfield  {journal} {\bibinfo  {journal} {Physical
  review B}\ }\textbf {\bibinfo {volume} {13}},\ \bibinfo {pages} {5188}
  (\bibinfo {year} {1976})}\BibitemShut {NoStop}%
\bibitem [{\citenamefont {Yu}\ \emph {et~al.}(2024)\citenamefont {Yu},
  \citenamefont {Herzog-Arbeitman}, \citenamefont {Wang}, \citenamefont
  {Vafek}, \citenamefont {Bernevig},\ and\ \citenamefont
  {Regnault}}]{yu2024fractional}%
  \BibitemOpen
  \bibfield  {author} {\bibinfo {author} {\bibfnamefont {J.}~\bibnamefont
  {Yu}}, \bibinfo {author} {\bibfnamefont {J.}~\bibnamefont
  {Herzog-Arbeitman}}, \bibinfo {author} {\bibfnamefont {M.}~\bibnamefont
  {Wang}}, \bibinfo {author} {\bibfnamefont {O.}~\bibnamefont {Vafek}},
  \bibinfo {author} {\bibfnamefont {B.~A.}\ \bibnamefont {Bernevig}},\ and\
  \bibinfo {author} {\bibfnamefont {N.}~\bibnamefont {Regnault}},\ }\href@noop
  {} {\bibfield  {journal} {\bibinfo  {journal} {Physical Review B}\ }\textbf
  {\bibinfo {volume} {109}},\ \bibinfo {pages} {045147} (\bibinfo {year}
  {2024})}\BibitemShut {NoStop}%
\bibitem [{\citenamefont {Jia}\ \emph {et~al.}(2024)\citenamefont {Jia},
  \citenamefont {Yu}, \citenamefont {Liu}, \citenamefont {Herzog-Arbeitman},
  \citenamefont {Qi}, \citenamefont {Pi}, \citenamefont {Regnault},
  \citenamefont {Weng}, \citenamefont {Bernevig},\ and\ \citenamefont
  {Wu}}]{jia2024moire}%
  \BibitemOpen
  \bibfield  {author} {\bibinfo {author} {\bibfnamefont {Y.}~\bibnamefont
  {Jia}}, \bibinfo {author} {\bibfnamefont {J.}~\bibnamefont {Yu}}, \bibinfo
  {author} {\bibfnamefont {J.}~\bibnamefont {Liu}}, \bibinfo {author}
  {\bibfnamefont {J.}~\bibnamefont {Herzog-Arbeitman}}, \bibinfo {author}
  {\bibfnamefont {Z.}~\bibnamefont {Qi}}, \bibinfo {author} {\bibfnamefont
  {H.}~\bibnamefont {Pi}}, \bibinfo {author} {\bibfnamefont {N.}~\bibnamefont
  {Regnault}}, \bibinfo {author} {\bibfnamefont {H.}~\bibnamefont {Weng}},
  \bibinfo {author} {\bibfnamefont {B.~A.}\ \bibnamefont {Bernevig}},\ and\
  \bibinfo {author} {\bibfnamefont {Q.}~\bibnamefont {Wu}},\ }\href@noop {}
  {\bibfield  {journal} {\bibinfo  {journal} {Physical Review B}\ }\textbf
  {\bibinfo {volume} {109}},\ \bibinfo {pages} {205121} (\bibinfo {year}
  {2024})}\BibitemShut {NoStop}%
\bibitem [{\citenamefont {Medvedeva}\ \emph {et~al.}(2017)\citenamefont
  {Medvedeva}, \citenamefont {Iskakov}, \citenamefont {Krien}, \citenamefont
  {Mazurenko},\ and\ \citenamefont {Lichtenstein}}]{medvedeva2017exact}%
  \BibitemOpen
  \bibfield  {author} {\bibinfo {author} {\bibfnamefont {D.}~\bibnamefont
  {Medvedeva}}, \bibinfo {author} {\bibfnamefont {S.}~\bibnamefont {Iskakov}},
  \bibinfo {author} {\bibfnamefont {F.}~\bibnamefont {Krien}}, \bibinfo
  {author} {\bibfnamefont {V.~V.}\ \bibnamefont {Mazurenko}},\ and\ \bibinfo
  {author} {\bibfnamefont {A.~I.}\ \bibnamefont {Lichtenstein}},\ }\href@noop
  {} {\bibfield  {journal} {\bibinfo  {journal} {Physical Review B}\ }\textbf
  {\bibinfo {volume} {96}},\ \bibinfo {pages} {235149} (\bibinfo {year}
  {2017})}\BibitemShut {NoStop}%
\bibitem [{\citenamefont {Lu}\ and\ \citenamefont
  {Haverkort}(2017)}]{Lu2017-ts}%
  \BibitemOpen
  \bibfield  {author} {\bibinfo {author} {\bibfnamefont {Y.}~\bibnamefont
  {Lu}}\ and\ \bibinfo {author} {\bibfnamefont {M.~W.}\ \bibnamefont
  {Haverkort}},\ }\href@noop {} {\bibfield  {journal} {\bibinfo  {journal}
  {Eur. Phys. J. Spec. Top.}\ }\textbf {\bibinfo {volume} {226}},\ \bibinfo
  {pages} {2549} (\bibinfo {year} {2017})}\BibitemShut {NoStop}%
\bibitem [{\citenamefont {Kotliar}\ and\ \citenamefont
  {Vollhardt}(2004)}]{Kotliar2004-pr}%
  \BibitemOpen
  \bibfield  {author} {\bibinfo {author} {\bibfnamefont {G.}~\bibnamefont
  {Kotliar}}\ and\ \bibinfo {author} {\bibfnamefont {D.}~\bibnamefont
  {Vollhardt}},\ }\href@noop {} {\bibfield  {journal} {\bibinfo  {journal}
  {Phys. Today}\ }\textbf {\bibinfo {volume} {57}},\ \bibinfo {pages} {53}
  (\bibinfo {year} {2004})}\BibitemShut {NoStop}%
\bibitem [{\citenamefont {Haule}(2007)}]{haule2007quantum}%
  \BibitemOpen
  \bibfield  {author} {\bibinfo {author} {\bibfnamefont {K.}~\bibnamefont
  {Haule}},\ }\href@noop {} {\bibfield  {journal} {\bibinfo  {journal}
  {Physical Review B—Condensed Matter and Materials Physics}\ }\textbf
  {\bibinfo {volume} {75}},\ \bibinfo {pages} {155113} (\bibinfo {year}
  {2007})}\BibitemShut {NoStop}%
\bibitem [{\citenamefont {Bauer}\ \emph {et~al.}(2011)\citenamefont {Bauer},
  \citenamefont {Carr}, \citenamefont {Evertz}, \citenamefont {Feiguin},
  \citenamefont {Freire}, \citenamefont {Fuchs}, \citenamefont {Gamper},
  \citenamefont {Gukelberger}, \citenamefont {Gull}, \citenamefont {Guertler},
  \citenamefont {Hehn}, \citenamefont {Igarashi}, \citenamefont {Isakov},
  \citenamefont {Koop}, \citenamefont {Ma}, \citenamefont {Mates},
  \citenamefont {Matsuo}, \citenamefont {Parcollet}, \citenamefont
  {Paw{\l}owski}, \citenamefont {Picon}, \citenamefont {Pollet}, \citenamefont
  {Santos}, \citenamefont {Scarola}, \citenamefont {Schollw{\"o}ck},
  \citenamefont {Silva}, \citenamefont {Surer}, \citenamefont {Todo},
  \citenamefont {Trebst}, \citenamefont {Troyer}, \citenamefont {Wall},
  \citenamefont {Werner},\ and\ \citenamefont {Wessel}}]{Bauer2011-ed}%
  \BibitemOpen
  \bibfield  {author} {\bibinfo {author} {\bibfnamefont {B.}~\bibnamefont
  {Bauer}}, \bibinfo {author} {\bibfnamefont {L.~D.}\ \bibnamefont {Carr}},
  \bibinfo {author} {\bibfnamefont {H.~G.}\ \bibnamefont {Evertz}}, \bibinfo
  {author} {\bibfnamefont {A.}~\bibnamefont {Feiguin}}, \bibinfo {author}
  {\bibfnamefont {J.}~\bibnamefont {Freire}}, \bibinfo {author} {\bibfnamefont
  {S.}~\bibnamefont {Fuchs}}, \bibinfo {author} {\bibfnamefont
  {L.}~\bibnamefont {Gamper}}, \bibinfo {author} {\bibfnamefont
  {J.}~\bibnamefont {Gukelberger}}, \bibinfo {author} {\bibfnamefont
  {E.}~\bibnamefont {Gull}}, \bibinfo {author} {\bibfnamefont {S.}~\bibnamefont
  {Guertler}}, \bibinfo {author} {\bibfnamefont {A.}~\bibnamefont {Hehn}},
  \bibinfo {author} {\bibfnamefont {R.}~\bibnamefont {Igarashi}}, \bibinfo
  {author} {\bibfnamefont {S.~V.}\ \bibnamefont {Isakov}}, \bibinfo {author}
  {\bibfnamefont {D.}~\bibnamefont {Koop}}, \bibinfo {author} {\bibfnamefont
  {P.~N.}\ \bibnamefont {Ma}}, \bibinfo {author} {\bibfnamefont
  {P.}~\bibnamefont {Mates}}, \bibinfo {author} {\bibfnamefont
  {H.}~\bibnamefont {Matsuo}}, \bibinfo {author} {\bibfnamefont
  {O.}~\bibnamefont {Parcollet}}, \bibinfo {author} {\bibfnamefont
  {G.}~\bibnamefont {Paw{\l}owski}}, \bibinfo {author} {\bibfnamefont {J.~D.}\
  \bibnamefont {Picon}}, \bibinfo {author} {\bibfnamefont {L.}~\bibnamefont
  {Pollet}}, \bibinfo {author} {\bibfnamefont {E.}~\bibnamefont {Santos}},
  \bibinfo {author} {\bibfnamefont {V.~W.}\ \bibnamefont {Scarola}}, \bibinfo
  {author} {\bibfnamefont {U.}~\bibnamefont {Schollw{\"o}ck}}, \bibinfo
  {author} {\bibfnamefont {C.}~\bibnamefont {Silva}}, \bibinfo {author}
  {\bibfnamefont {B.}~\bibnamefont {Surer}}, \bibinfo {author} {\bibfnamefont
  {S.}~\bibnamefont {Todo}}, \bibinfo {author} {\bibfnamefont {S.}~\bibnamefont
  {Trebst}}, \bibinfo {author} {\bibfnamefont {M.}~\bibnamefont {Troyer}},
  \bibinfo {author} {\bibfnamefont {M.~L.}\ \bibnamefont {Wall}}, \bibinfo
  {author} {\bibfnamefont {P.}~\bibnamefont {Werner}},\ and\ \bibinfo {author}
  {\bibfnamefont {S.}~\bibnamefont {Wessel}},\ }\href@noop {} {\bibfield
  {journal} {\bibinfo  {journal} {J. Stat. Mech.}\ }\textbf {\bibinfo {volume}
  {2011}},\ \bibinfo {pages} {P05001} (\bibinfo {year} {2011})}\BibitemShut
  {NoStop}%
\end{thebibliography}%

\clearpage
\newpage

\onecolumngrid

\section{Supporting Information}

\renewcommand{\figurename}{Figure S\!\!}
\renewcommand{\tablename}{Table S\!\!}
\setcounter{figure}{0}  


\begin{figure*}[h!]
\includegraphics[width=0.8\textwidth]{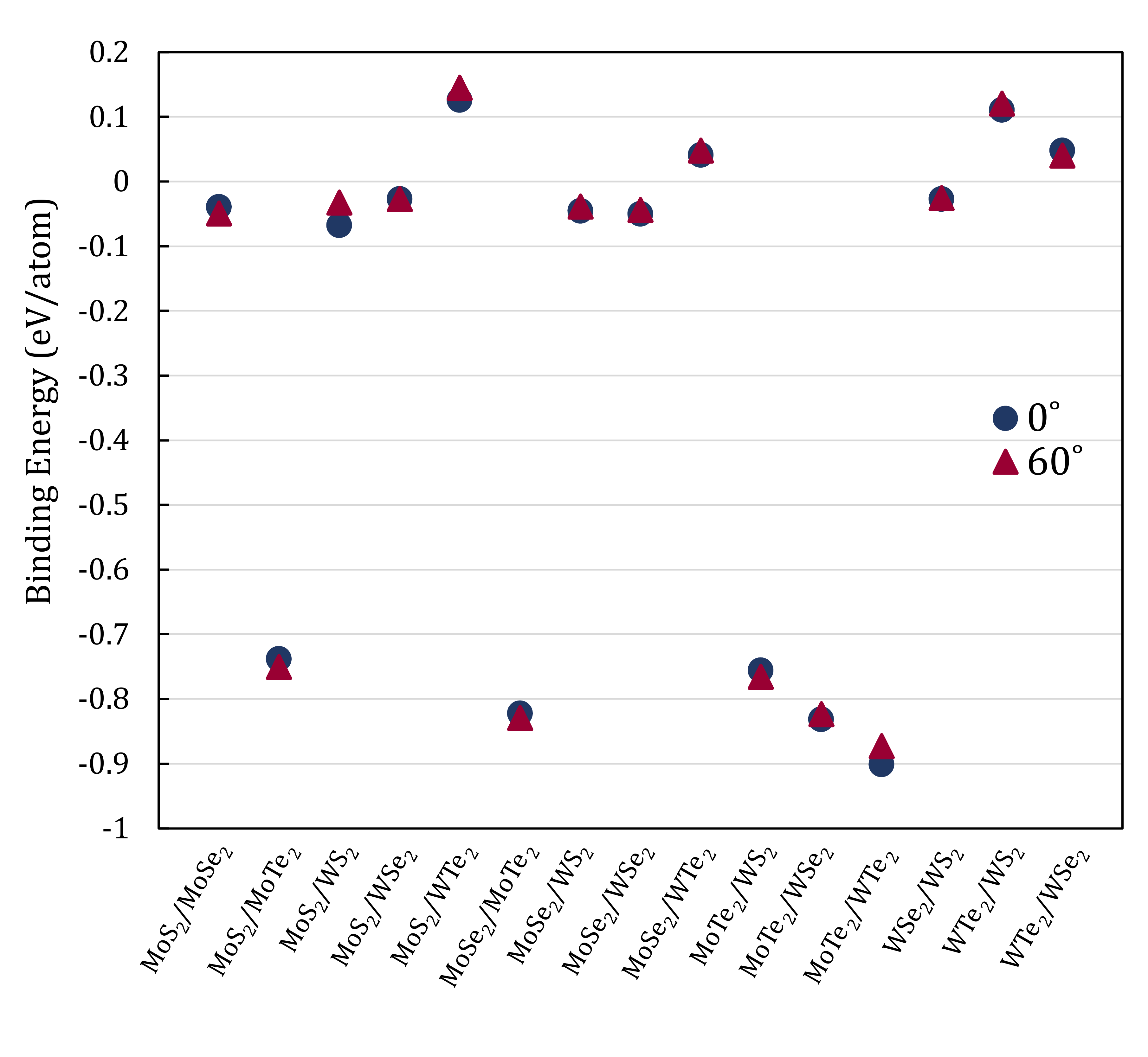}
\centering
\caption{\footnotesize Binding energy are shown for 15 bilayer TMDCs heterostructures with two types of stacking, either 0 or 60 degrees of shifting.}
\label{fig:S1}
\end{figure*}

\begin{figure*}[h!]
\includegraphics[width=0.8\textwidth]{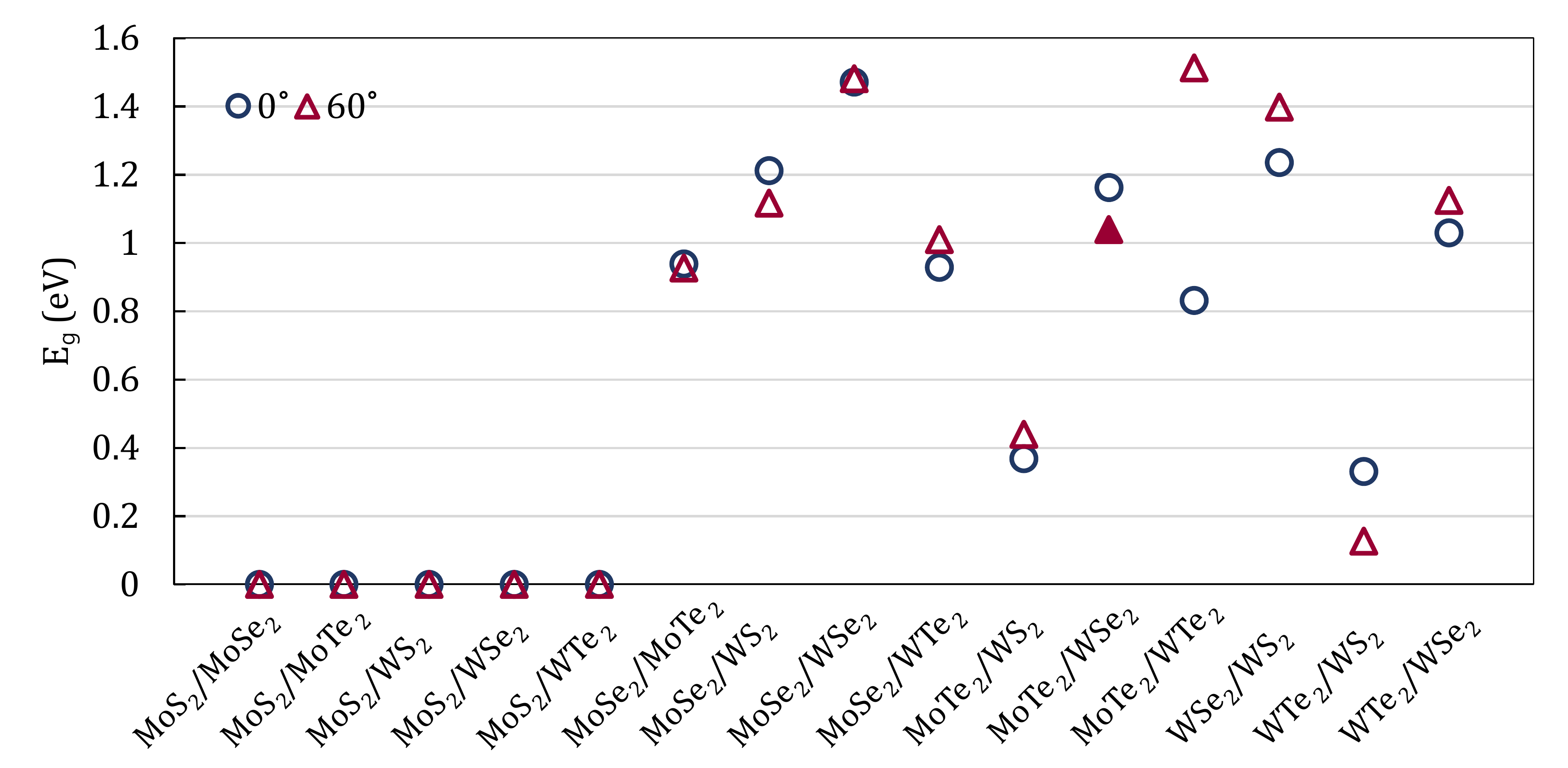}
\centering
\caption{\footnotesize The E$_g$ of 15 bilayer TMDCs heterostructures compared with two types of stacking, either 0 or 60 degrees of shifting. Hollow signs represent indirect E$_g$ while filled sign stands for direct E$_g$. Compared to this plot where 1T-MoS$_2$ was applied, 1H-MoS$_2$ was applied in Figure 2.}
\label{fig:S2}
\end{figure*}

\begin{figure*}
\includegraphics[width=0.6\textwidth]{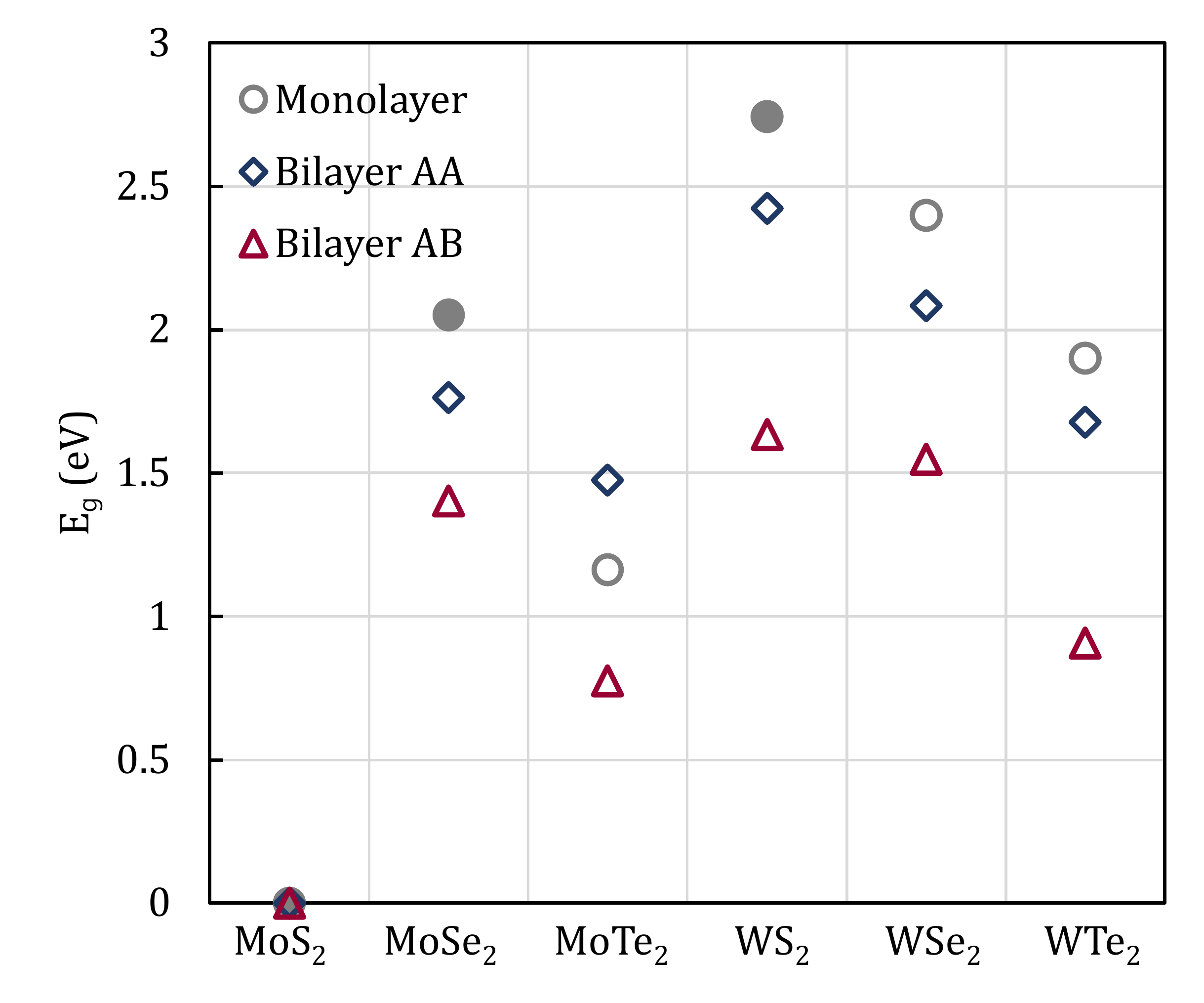}
\centering
\caption{\footnotesize The E$_g$ are shown of 6 fully optimized TMDCs in forms of monolayer, A-A stacking bilayer, and A-B stacking bilayer. The MoS$_2$ in this plot belongs to 1T instead of 1H. Hollow signs represent indirect E$_g$ while filled signs stand for direct E$_g$.}
\label{fig:S3}
\end{figure*}

\begin{figure*}
\includegraphics[width=0.8\textwidth]{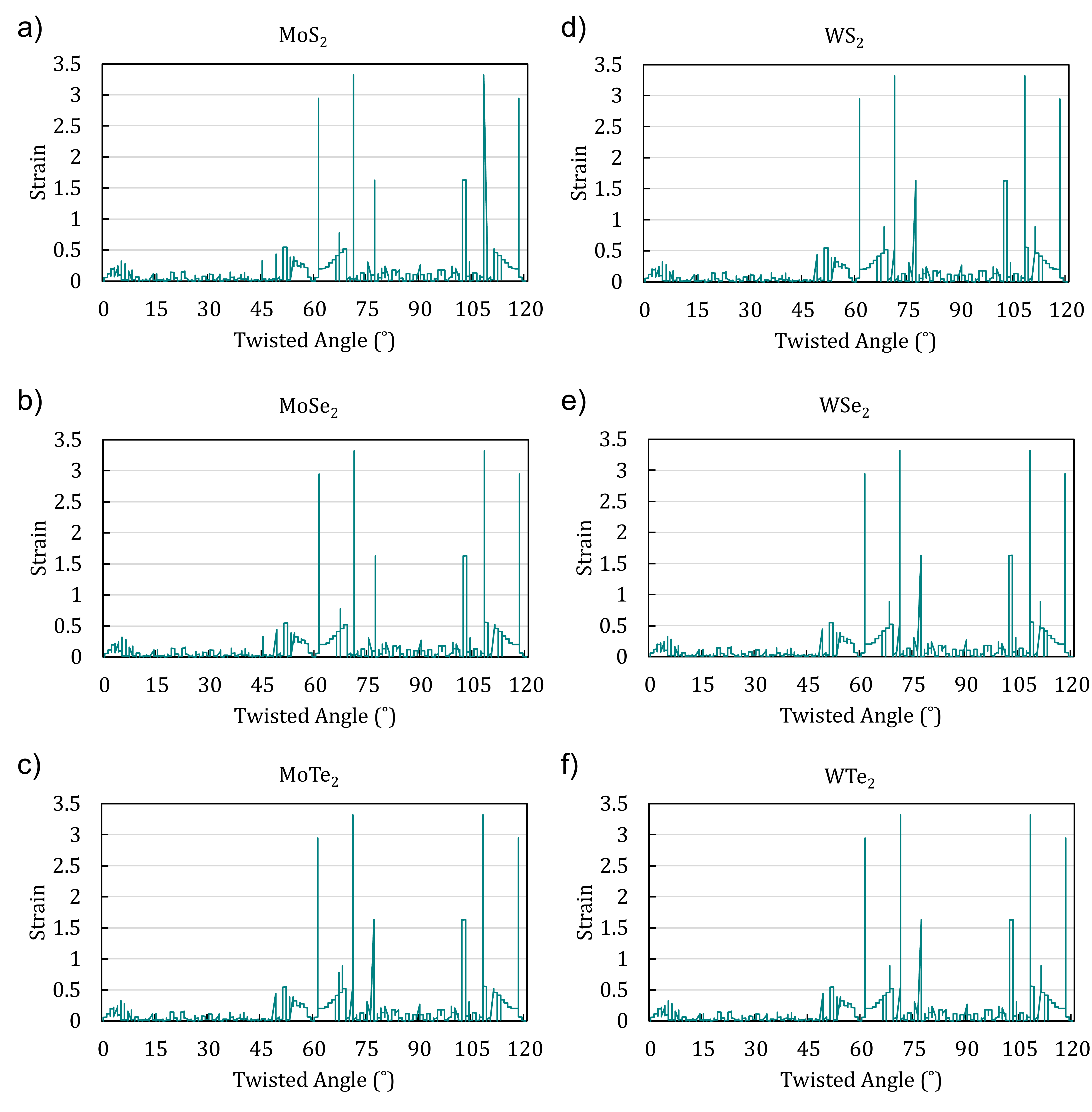}
\centering
\caption{\footnotesize The overall fluctuation of strain according to the twisted angles of 1H bilayers are shown for the cases of (a) MoS$_2$, (b) MoSe$_2$, (c) MoTe$_2$, (d) WS$_2$, (e) WSe$_2$, and (f) WTe$_2$.}
\label{fig:S4}
\end{figure*}

\begin{figure*}
\includegraphics[width=0.8\textwidth]{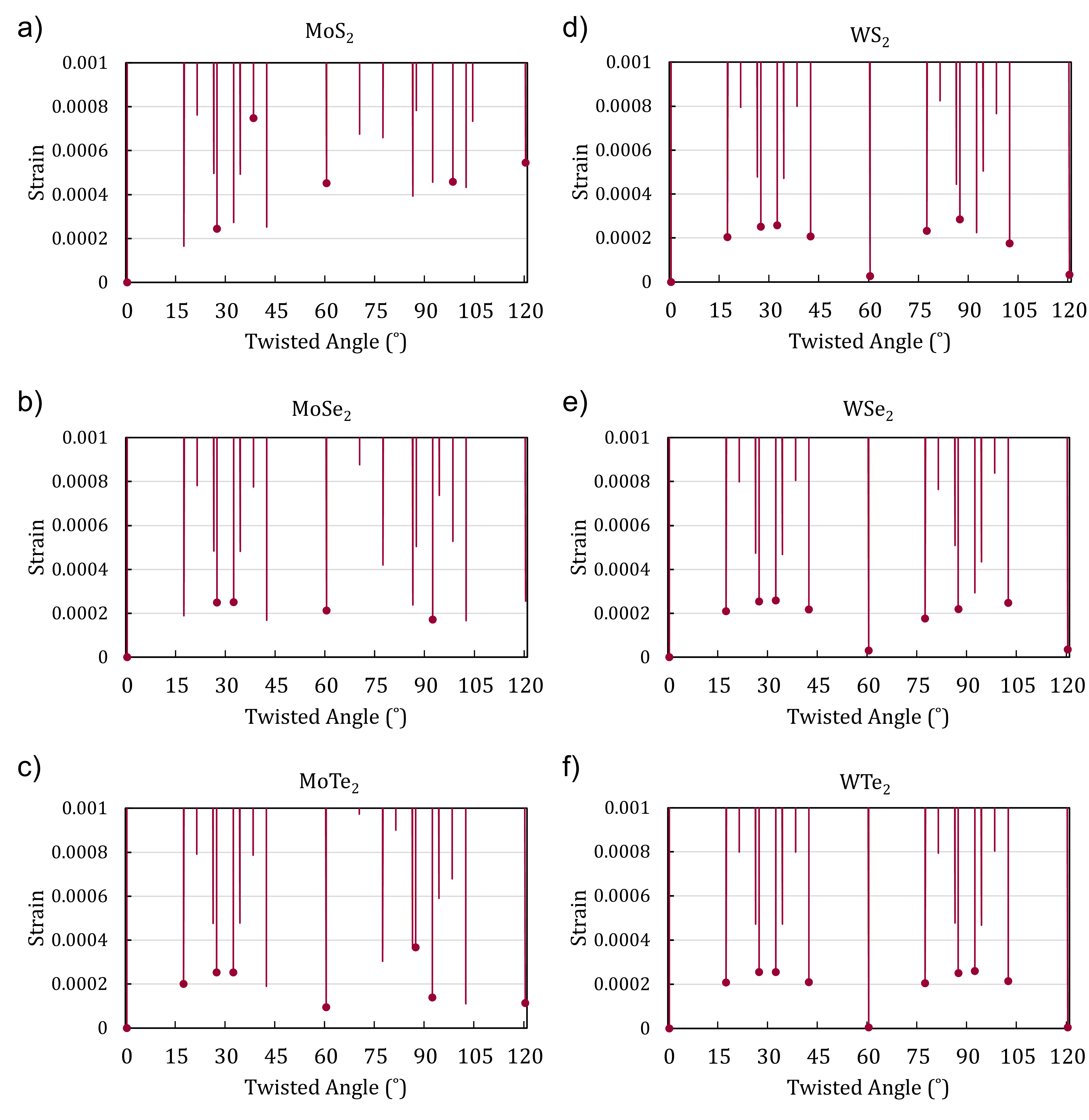}
\centering
\caption{\footnotesize The zoom-in fluctuation of strain below 0.001 according to the twisted angles of 1H bilayers in the cases of (a) MoS$_2$, (b) MoSe$_2$, (c) MoTe$_2$, (d) WS$_2$, (e) WSe$_2$, and (f) WTe$_2$. The low-strain twisted TMDCs bilayers being symmetrized and further computed in this work are dotted in the plots.}
\label{fig:S5}
\end{figure*}

\begin{figure*}
\includegraphics[width=0.8\textwidth]{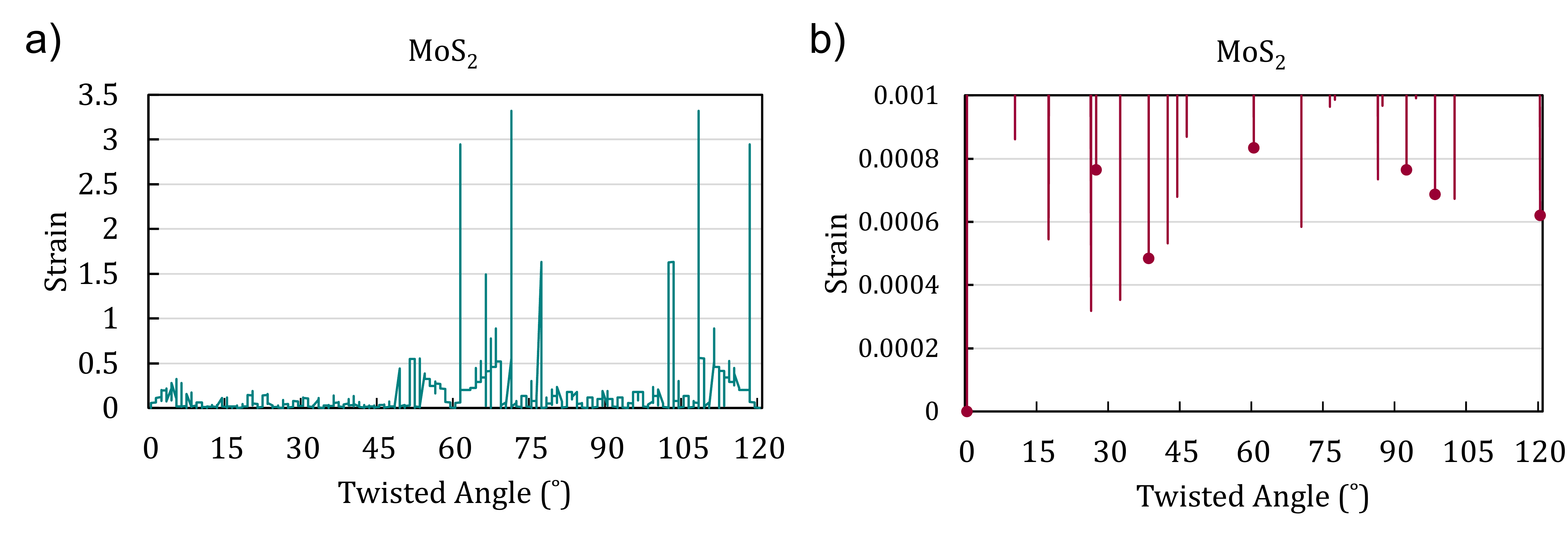}
\centering
\caption{\footnotesize (a) The overall fluctuation of strain according to the twisted angles of bilayers in the cases of 1T-MoS$_2$. (b) The zoom-in fluctuation of strain below 0.001 according to the twisted angles of bilayers in the cases of 1T-MoS$_2$. The low-strain twisted TMDCs bilayers being symmetrized and further computed in this work are dotted in the plots.}
\label{fig:S6}
\end{figure*}

\begin{figure*}
\includegraphics[width=\textwidth]{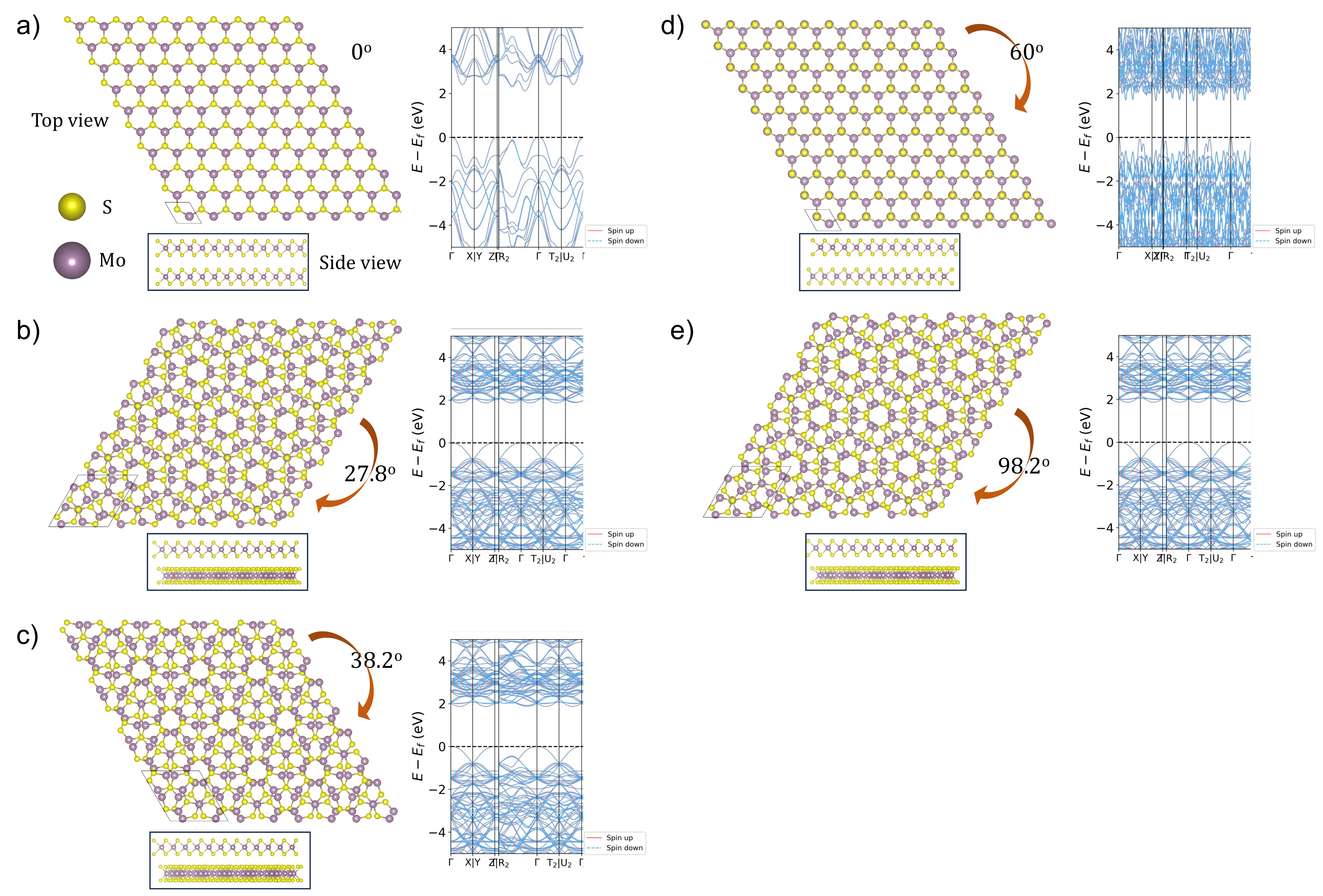}
\centering
\caption{\footnotesize The schematic geometries and band structures of bilayer 1H-MoS$_2$ with the twisted angles of (a) 0\textdegree~, (b) 27.8\textdegree~, (c) 38.2\textdegree~, (d) 60\textdegree~, and (e) 98.2\textdegree~. These include the top view and the side view of geometries (bottom left).}
\label{fig:S7}
\end{figure*}

\begin{figure*}
\includegraphics[width=\textwidth]{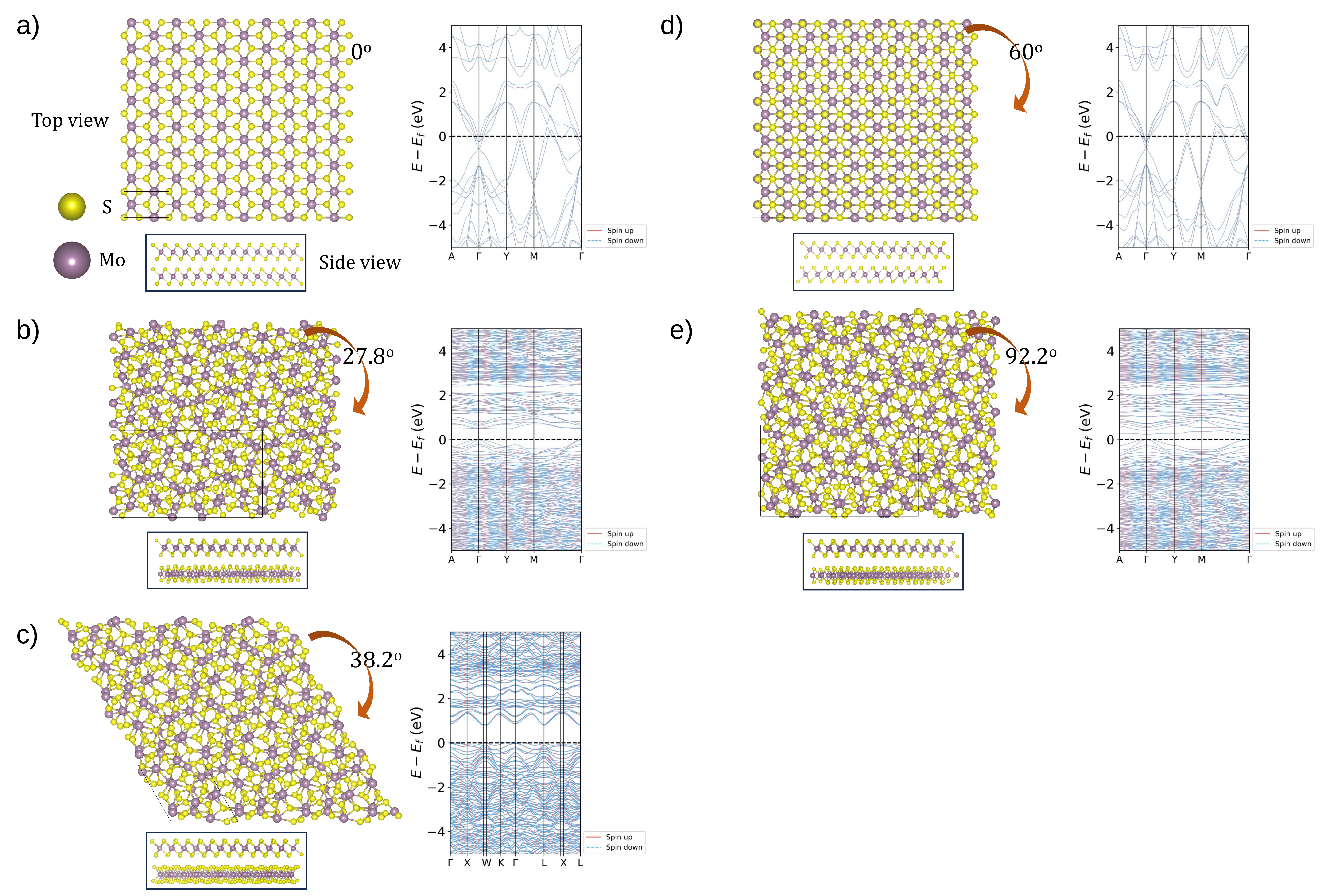}
\centering
\caption{\footnotesize The schematic geometries and band structures of bilayer 1T-MoS$_2$ with the twisted angles of (a) 0\textdegree~, (b) 27.8\textdegree~, (c) 38.2\textdegree~, (d) 60\textdegree~, and (e) 92.2\textdegree~. These include the top view and the side view of geometries (bottom left).}
\label{fig:S8}
\end{figure*}

\begin{figure*}
\includegraphics[width=\textwidth]{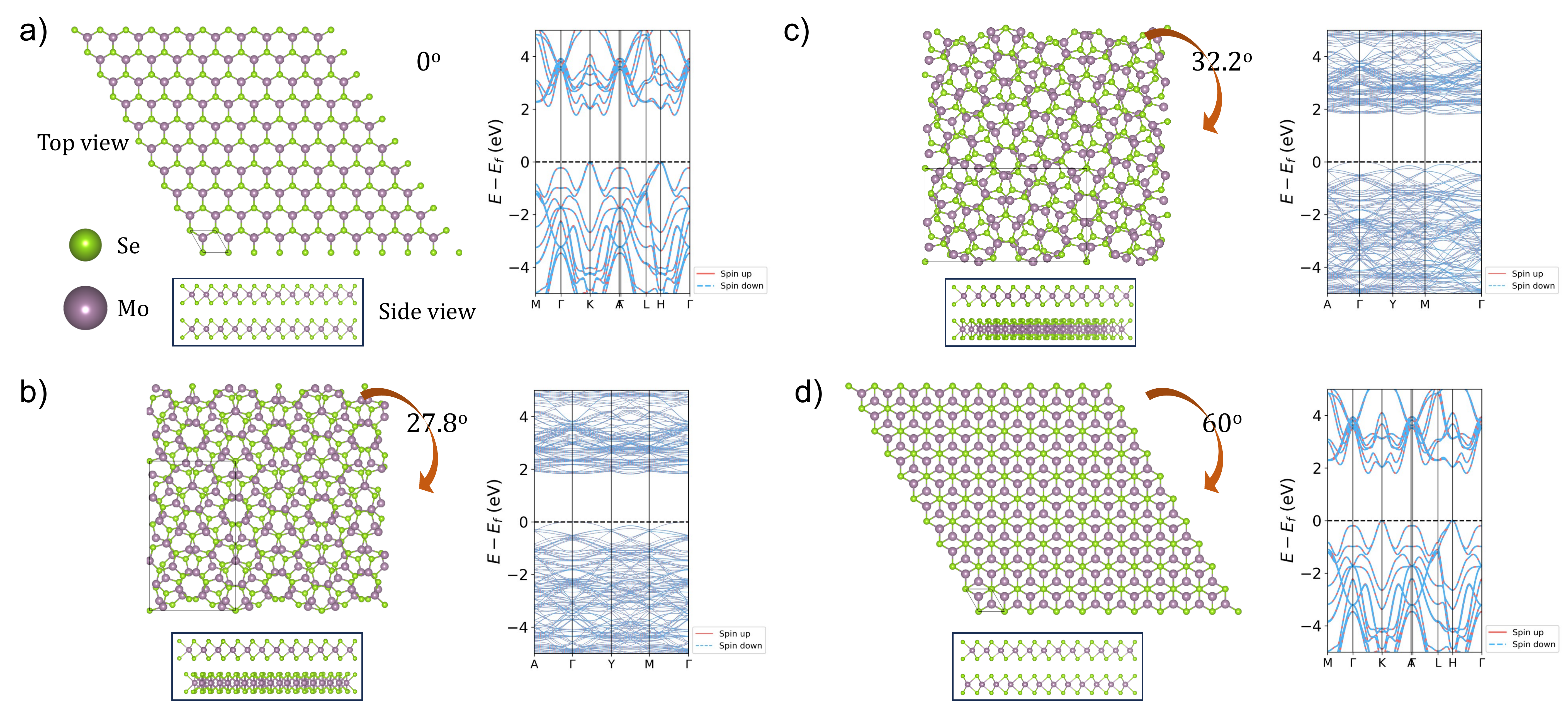}
\centering
\caption{\footnotesize The schematic geometries and band structures of bilayer MoSe$_2$ with the twisted angles of (a) 0\textdegree~, (b) 27.8\textdegree~, (c) 32.2\textdegree~, and (d) 60\textdegree~. These include the top view and the side view of geometries (bottom left).}
\label{fig:S9}
\end{figure*}

\begin{figure*}
\includegraphics[width=\textwidth]{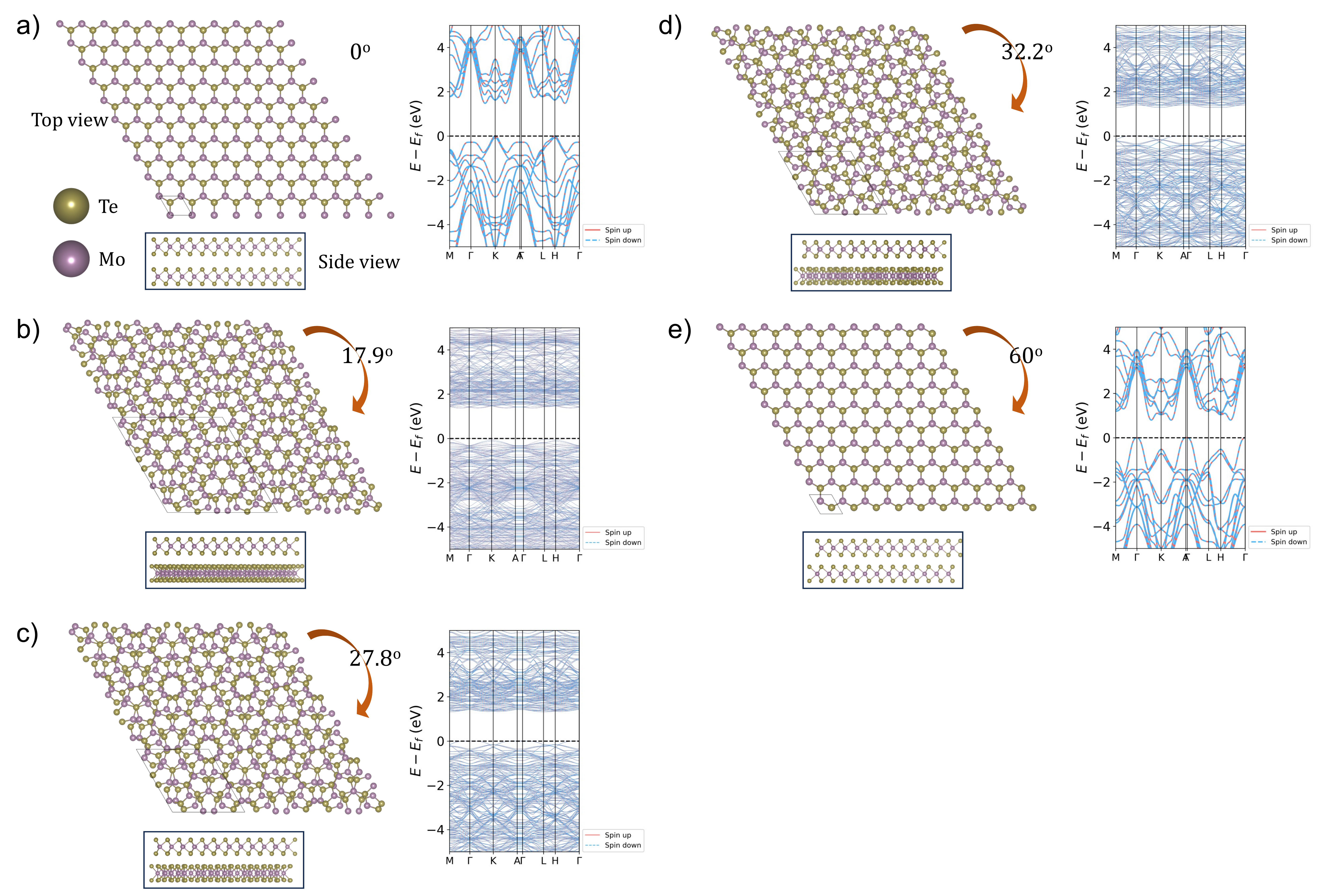}
\centering
\caption{\footnotesize The schematic geometries and band structures of bilayer MoTe$_2$ with the twisted angles of (a) 0\textdegree~, (b) 17.9\textdegree~, (c) 27.8\textdegree~, (d) 32.2\textdegree~, and (e) 60\textdegree~. These include the top view and the side view of geometries (bottom left).}
\label{fig:S10}
\end{figure*}

\begin{figure*}
\includegraphics[width=\textwidth]{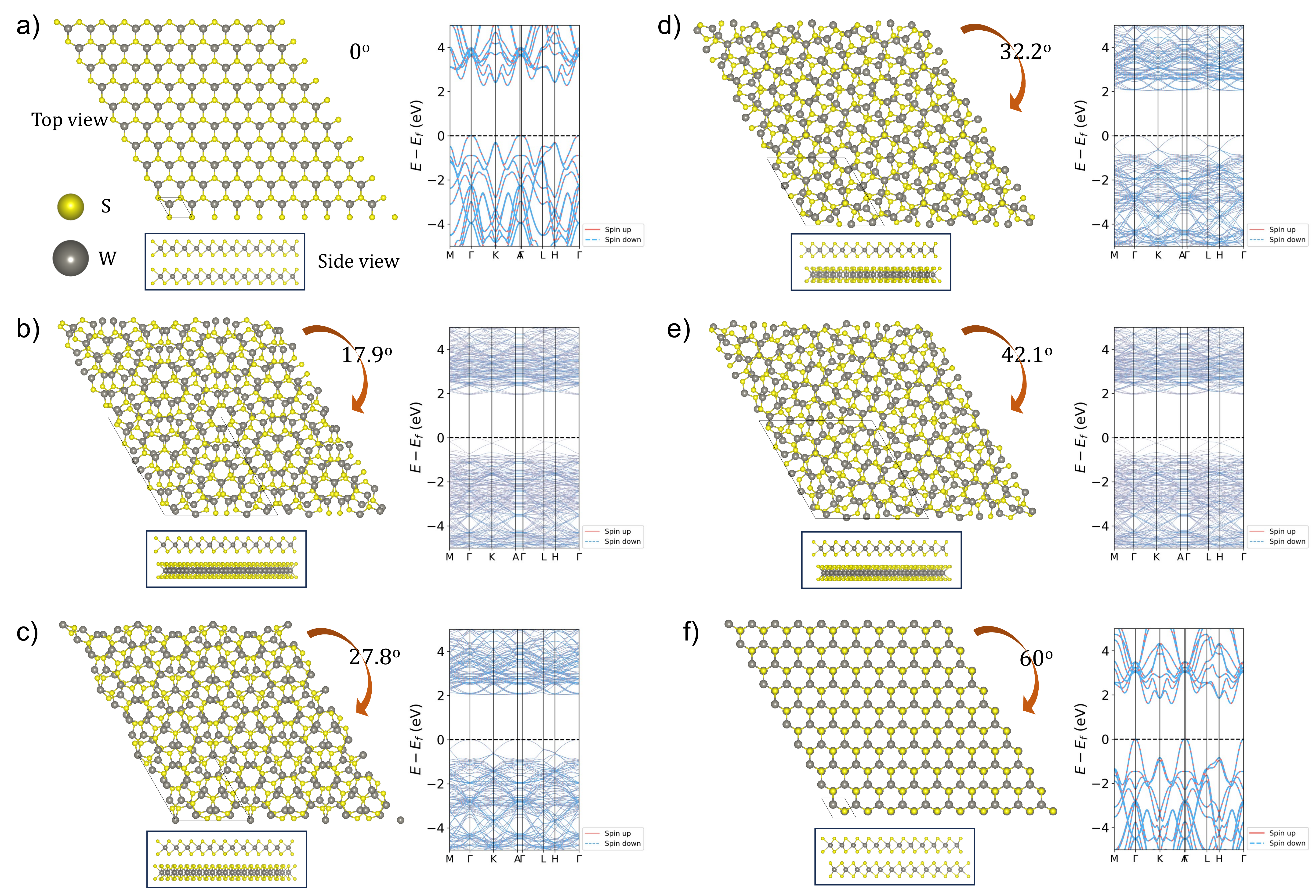}
\centering
\caption{\footnotesize The schematic geometries and band structures of bilayer WS$_2$ with the twisted angles of (a) 0\textdegree~, (b) 17.9\textdegree~, (c) 27.8\textdegree~, (d) 32.2\textdegree~, (e) 42.1\textdegree~, and (f) 60\textdegree~. These include the top view and the side view of geometries (bottom left).}
\label{fig:S11}
\end{figure*}

\begin{figure*}
\includegraphics[width=0.9\textwidth]{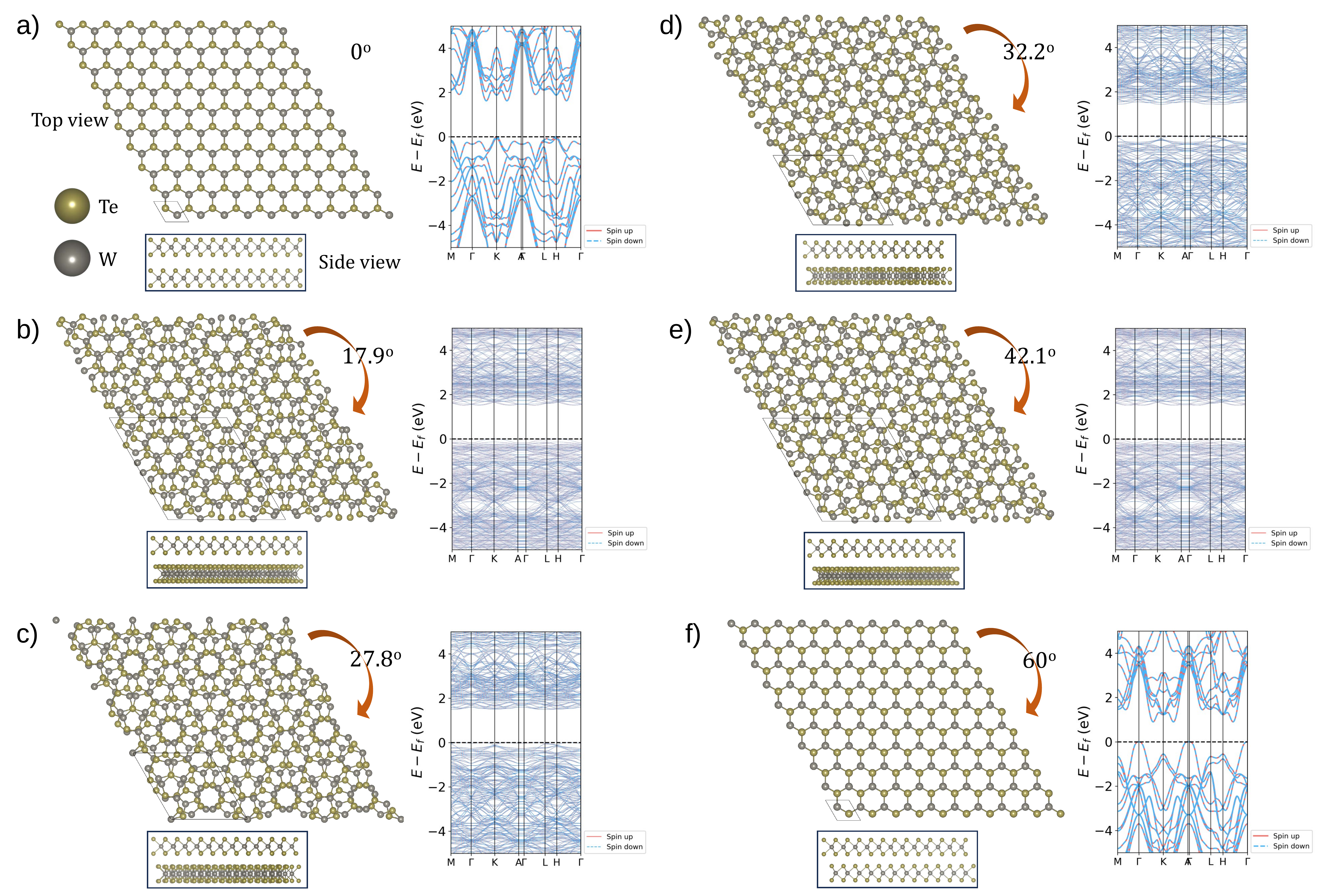}
\centering
\caption{\footnotesize The schematic geometries and band structures of bilayer WTe$_2$ with the twisted angles of (a) 0\textdegree~, (b) 17.9\textdegree~, (c) 27.8\textdegree~, (d) 32.2\textdegree~, (e) 42.1\textdegree~, and (f) 60\textdegree~. These include the top view and the side view of geometries (bottom left).}
\label{fig:S12}
\end{figure*}

\begin{figure}[h]
    \centering
    \includegraphics[width=\textwidth]{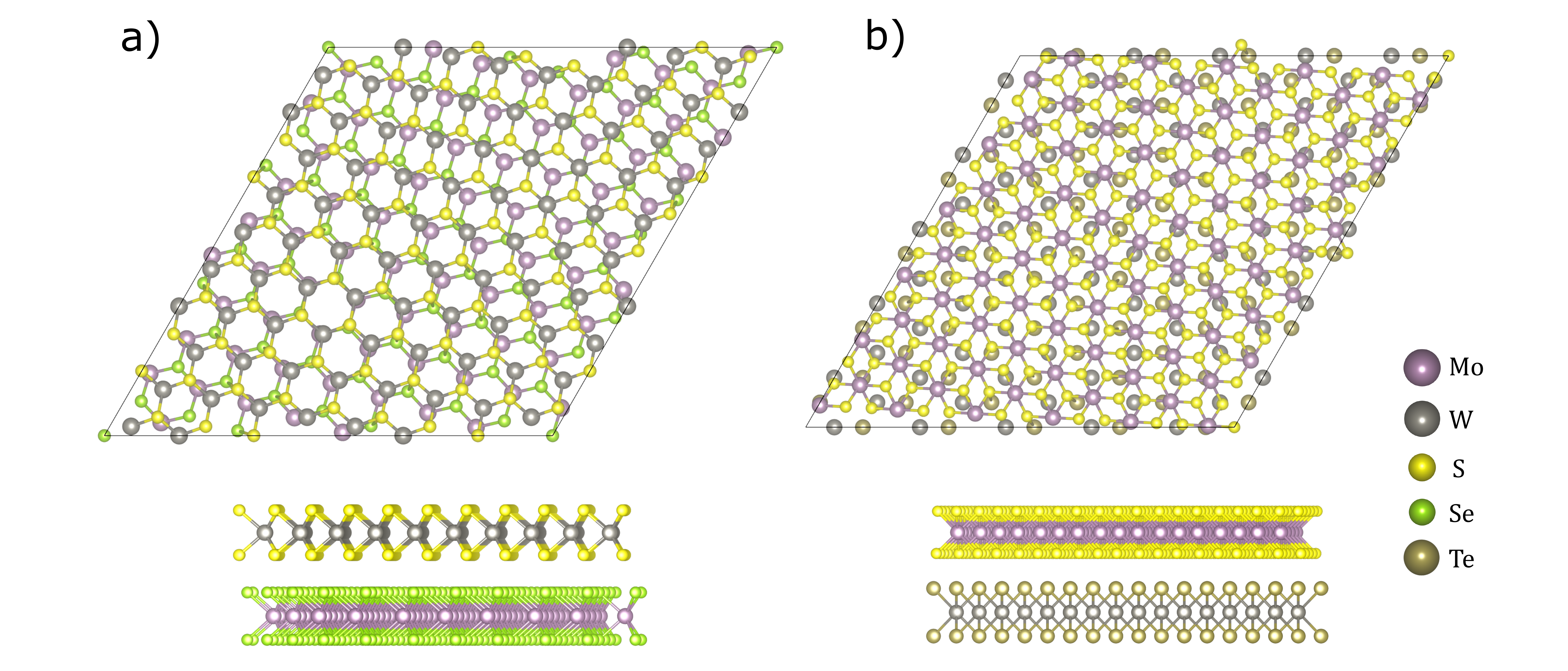}
    \caption{ The schematic geometries of bilayer (a) MoSe$_2$/WS$_2$ at a twisted angle of  6.1\textdegree~, and (b) WTe$_2$/MoS$_2$ at a twisted angle of 3.0\textdegree~. These include the top view (top) and the side view (bottom) of geometries.}
    \label{fig:S13}
\end{figure}

\begin{figure}[h]
    \centering
    \includegraphics[width=\textwidth]{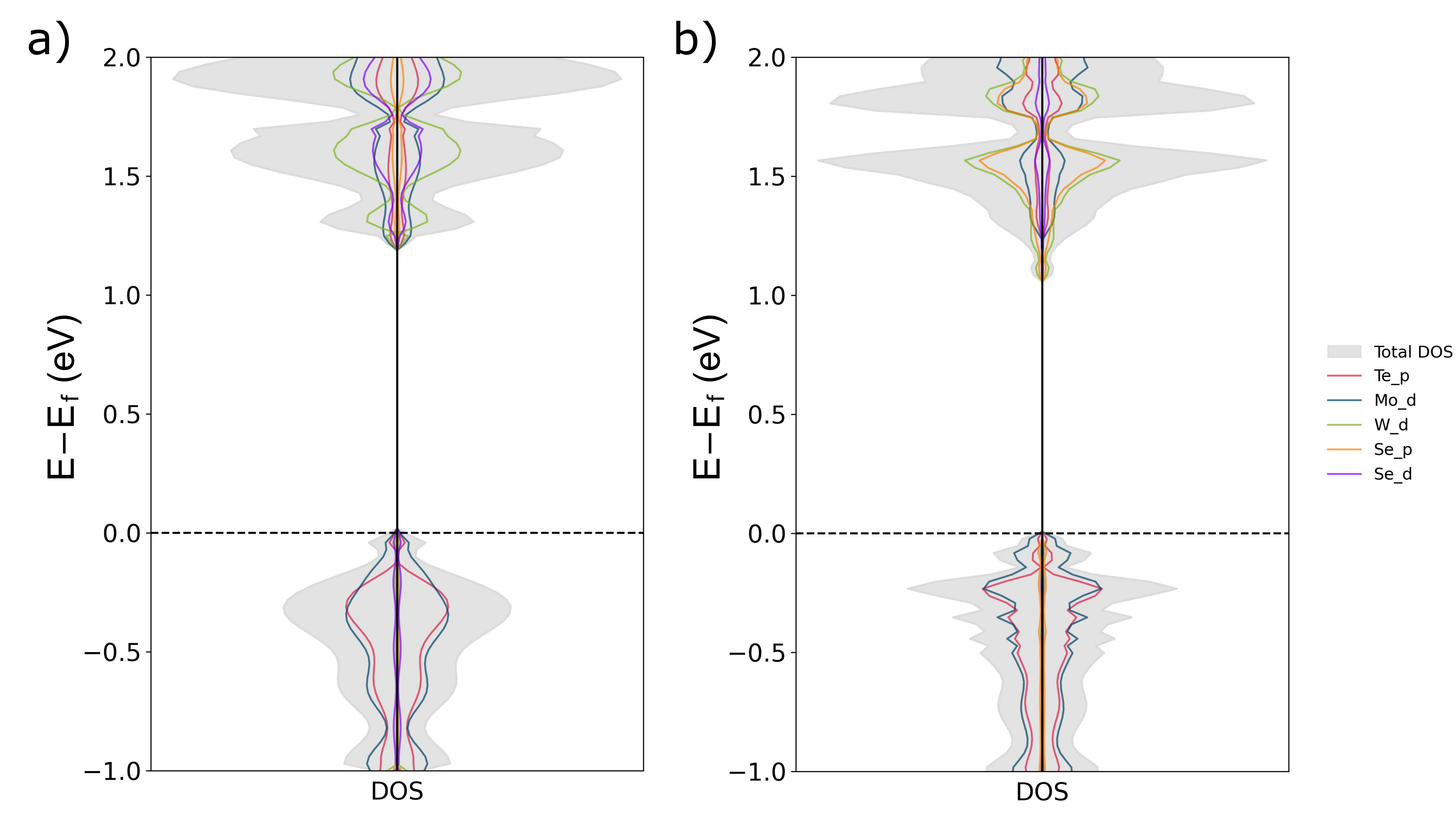}
    \caption{ The orbital projecting density of states of bilayer MoTe$_2$/WSe$_2$ at a stacking angle of (a) 0\textdegree~and (b) 60\textdegree~. The corresponding influential orbitals of each element are shown.}
    \label{fig:S14}
\end{figure}

\end{document}